\documentclass[aps,twocolumn,pra,superscriptaddress]{revtex4-2}
\usepackage{amsfonts}
\usepackage[final,hiresbb]{graphicx}
\usepackage{amsmath}
\usepackage{amssymb}
\usepackage{amstext}
\usepackage{color}
\usepackage{braket}

\usepackage{subfigure}
\usepackage{appendix}

\definecolor{Green}{RGB}{0,204,102}
\definecolor{Purple}{RGB}{102,0,255}
\definecolor{Blue}{RGB}{51,153,255}
\definecolor{Red}{RGB}{255,010,010}
\definecolor{Orange}{RGB}{255,192,0}

\newcommand\numberthis{\addtocounter{equation}{1}\tag{\theequation}}

\newcommand{\bfr}{{\bf r}_{\scriptscriptstyle\perp}}
\newcommand{\vecr}{\vec r}
\newcommand{\vecu}{\vec u}

\newcommand{\uLC}{\vec u_{\rm \scriptscriptstyle v}}

\newcommand{\pbg}{{\psi_{bg}}}

\newcommand{\pLC}{\psi_{\rm \scriptscriptstyle v}}

\newcommand{\gs}{{\nabla_\perp}}

\newcommand{\F}{{\bf F}}
\newcommand{\R}{{\bf R}}
\newcommand{\V}{{\bf V}}

\newcommand{\Vsq}{{\bf V}^2}

\newcommand{\FLC}{{\bf F}_{\rm \scriptscriptstyle v}}
\newcommand{\FLCinv}{{\bf F}_{\rm \scriptscriptstyle v}^{-1}}
\newcommand{\FLCinvtrans}{{\bf F}_{\rm \scriptscriptstyle v}^{-T}}
\newcommand{\RLC}{{\bf R}_{\rm \scriptscriptstyle v}}
\newcommand{\ULC}{{\bf U}_{\rm \scriptscriptstyle v}}
\newcommand{\VLC}{{\bf V}_{\rm \scriptscriptstyle v}}
\newcommand{\VLCsq}{{\bf V}_{\rm \scriptscriptstyle v}^2}
\newcommand{\JLC}{{J}_{\rm \scriptscriptstyle v}}
\newcommand{\Phibg}{{\boldsymbol \Phi}_{\rm \scriptscriptstyle bg}}

\newcommand{\pauli}{{\boldsymbol \sigma}_{0}}

\begin{document}
	\title{Hydrodynamics of noncircular vortices in beams of light and other two-dimensional fluids}
	%Why and How Tilt Affects Vortex Dynamics in Beams of Light and Other Two-Dimensional Fluids
	
	\author{Jasmine M. Andersen}
	\affiliation{Department of Physics and Astronomy, University of Denver, Denver, Colorado 80208, USA}
	
	\author{Andrew A. Voitiv}
	\affiliation{Department of Physics and Astronomy, University of Denver, Denver, Colorado 80208, USA}
	
	\author{Mark E. Siemens}
	\email{Mark.Siemens@du.edu}
	\affiliation{Department of Physics and Astronomy, University of Denver, Denver, Colorado 80208, USA}
	\author{Mark T. Lusk}
	\email{mlusk@mines.edu}
	\affiliation{Department of Physics, Colorado School of Mines, Golden, Colorado 80401, USA}

	\begin{abstract}
		The motion of noncircular two-dimensional vortices is shown to depend on a form of coupling between vortex ellipticity and the gradient of fluid density. The approach is based on the perspective that an elliptic vortex can be described as the projection of a virtual construct, a circular vortex with a symmetry axis that is tilted with respect to the direction of propagation. The resulting kinetic equation offers insights into how tilt and vortex velocity coevolve in few-body nonequilibrium settings such as vortex pair nucleation and annihilation. The model is developed and applied in association with optical vortices, and optical experiments are used to verify its predictive power. It is valid for quantum fluids and classical hydrodynamics settings as well.
	\end{abstract}
	\maketitle

	%%%%%%%%%%%%%%%%%%%%%%%%%%%%%%%%%%%%%%%%%%%%%%%%%%
	\section{Introduction}
	%%%%%%%%%%%%%%%%%%%%%%%%%%%%%%%%%%%%%%%%%%%%%%%%%%

%Knots and braids~\cite{Dennis2003, Leach2005}

%Fractional charge~\cite{Kotlyar2017}

%Poynting vector along z-axis.

The need to predict vortex motion in two-dimensional fluids is ubiquitous in disciplines ranging from weather prediction to quantum information science. There is a straightforward way of doing this for inviscid, incompressible media because vortices move with the underlying fluid.  The same is not true for fluids with a potential energy that depends on local density though. Although vortices are still constrained to move within bulk velocity contours, the contours themselves can become highly distorted, allowing vortices to move relative to the background fluid. Similar to the way in which interface and dislocation motion is described~\cite{Mullins1964,Knowles1990,Hirth1998}, a kinetic equation for vortices is then needed for both predictive power and physical insight. 

In settings for which such barotropic fluids move in a circular path about isolated vortices, a vortex kinetic equation has been derived~\cite{Nilsen2006,kivshar1998} and successfully applied~\cite{Jezek2008, dosSantos2016, Groszek2018}. The equation includes a correction term that accounts for the influence of fluid density gradients on vortex motion. While this is an important advance, most nonequilibrium vortex dynamics exhibit a distinctly elliptical shape that itself evolves as vortices move~\cite{Bershader1995,Leweke2016}. This is particularly evident when vortices of opposite rotation are nucleated or annihilated~\cite{kida1994, Eldredge2002}. In general, even accounting for density gradients results in incorrect predictions for the motion of vortices that exhibit ellipticity.

An ideal setting in which to study inviscid vortex dynamics is within coherent beams of light~\cite{Allen1992}. There, Maxwell's equations are a formal analogy to a free-space, two-dimensional Schr\"odinger equation~\cite{Lax1975} in which the direction of the time-averaged Poynting vector is viewed as a temporal axis for the transverse plane. A Madelung transformation~\cite{Madelung1927} re-casts the bulk dynamics as an Euler equation, allowing the interpretation of light as an inviscid, compressible, two-dimensional fluid. This hydrodynamic interpretation for linear optical propagation has been tested by tracking multiple unit-charge optical vortices as they move in a laser beam. While simple configurations of vortices with the same charge looked promising for a hydrodynamics interpretation~\cite{Rozas1999}, configurations of vortices that include opposite charges inevitably lead to vortex deformation and emergent ellipticity, which has not been accounted for in the hydrodynamic model. Notably, this ellipticity is observed in the plane transverse to laser propagation, in contrast to arbitrary ellipticity from rotating the measurement plane. 

For isolated vortex pairs of like and unlike charge, ellipticity can be analytically characterized~\cite{Singh2003, Bekshaev2003, Zhao2017, Kotlyar2017} along with expressions for vortex trajectories~\cite{Indebetouw1993,ChenRoux2008}, but these do not give physical insight into to why vortices move as they do. For such cases in which vortex ellipticity emerges, previously derived kinetic equations that assume canonical vortex shape cannot describe the dynamics ~\cite{Nilsen2006,kivshar1998,Groszek2018}. Therefore, it is clear that a hydrodynamic understanding of optical vortex motion must account for vortex ellipticity; what is not clear is how to do so.

In this paper, two-dimensional, elliptical vortices are treated as the projection of a circular vortex associated with an  axis of symmetry that is tilted with respect to the propagation direction of the beam. This formalism allows ellipticity to be described in terms of two angles, polar tilt and azimuthal orientation, instead of the orientation and aspect ratio of the actual ellipse. It simplifies both analysis and interpretation and amounts to describing an ellipse as the shadow of tilted disk, as shown in Fig. \ref{zprime_projection}. The resulting vortex kinetic equation can subsequently be used to predict the trajectories of elliptical vortices. Vortex ellipticity is described in terms of polar tilt and azimuthal orientation, and the kinetic equation shows that these angles are coupled to the gradient in fluid density. It applies to any two-dimensional, inviscid fluid including those with non-linear interaction terms that typify quantum fluids.
 
This theory provides a means of predicting elliptical vortex motion in a laser beam and offers a hydrodynamic interpretation of the dynamics. It is accompanied by methods for generating and measuring various optical vortex tilts in the laboratory and quantitatively validates the vortex kinetic theory with two experiments which demonstrate how vortex tilt affects its motion: one in which a vortex moves across the shoulder of a Gaussian beam and the other in which two vortices undergo an annihilation event.

%%%%%%%%%%%%%%%%%%%%%%%%%%%%%%%%%%%%%%%%%%%%%%%%%%
\section{Paraxial/Schr\"odinger Tilted Vortices}
%%%%%%%%%%%%%%%%%%%%%%%%%%%%%%%%%%%%%%%%%%%%%%%%%%

\subsection{Theoretical Formulation}%Tilted_Single_Vortex_Gaussian_110319_Euler.nb
% Vortex_Tilt_and_Velocity_122320.nb
\label{sec:tiltformalism}

Under a paraxial approximation for the monochromatic electromagnetic vector potential, ${\bf A}(\bfr , t) = {\bf e}_x A_0\psi(\bfr,z)e^{i(k z - \omega t)}$, electrodynamics are governed by a two-dimensional Schr\"odinger equation~\cite{Lax1975}:
\begin{equation}\label{paraxial}
	i \partial_z \psi = -\frac{1}{2k} \nabla^2_\perp \psi + {\cal V}\psi.
\end{equation}
Here $\cal V$ is an arbitrary linear operator that can even be a nonlinear function of $\psi$, allowing the results which follow to be applied to both nonlinear optics and quantum fluid settings. Two characteristic lengths, which vary with the application,  are used to nondimensionalize the transverse and axial positions, so that the resulting nondimensional equation is 
\begin{equation}\label{paraxial_ND}
	i \partial_z \psi = -\frac{1}{2} \nabla^2_\perp \psi + {\cal V}\psi .
\end{equation}
Here and in the following theoretical development, all positions are taken to be nondimensional.

Without loss of generality, the paraxial field in the vicinity of a vortex, $\psi$, can be written as the product of a field associated with a linear-core vortex and a background field such that
\begin{equation}\label{eq:ppbgpv}
    \psi= \pbg \psi_{\scriptscriptstyle v}
\end{equation}
\noindent where $\psi_{\scriptscriptstyle v}$ and $\pbg$ are the vortex and background field, respectively. For instance, a Laguerre-Gaussian beam with $\ell=1$, $p=0$, beam waist $w_0$, and characteristic transverse length $w_0/\sqrt{2}$ has a Gaussian background field and a vortex with circular symmetry
\begin{equation}\label{LC}
	\pbg(x,y,0) = \frac{1}{\sqrt{\pi}} e^{-(x^2+y^2)/2}, \quad \psi_{\scriptscriptstyle v} = x + i y .
\end{equation}
Both here and in general, an accurate vortex description makes it possible to distill out the locally singular behavior of a paraxial vortex field. In the case of elliptical vortices, the background field can only be determined if the form of $\psi_v$ includes its ellipticity.

\begin{figure}[t]
	\begin{center}
		\includegraphics[width=0.85\linewidth]{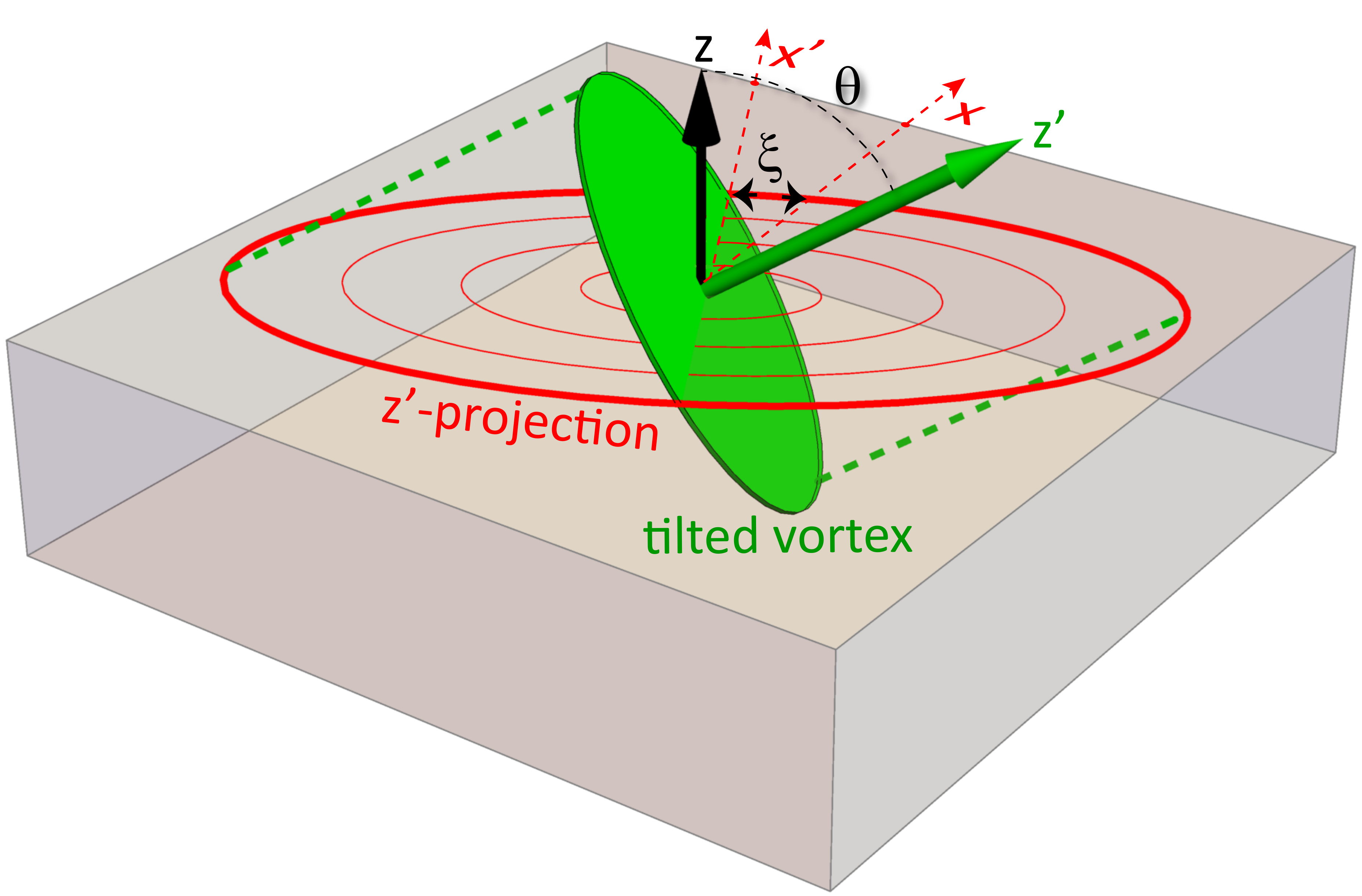}
	\end{center}
	\caption{ \emph{Tilted vortex perspective}. A 2D elliptical vortex in the transverse plane can be viewed as the projection, along the tilt axis, of a 3D vortex. The ellipticity can then be described by an azimuthal orientation angle $\xi$ and a polar lean $\theta$.} 
	\label{zprime_projection}
\end{figure}

In the following, we develop a formalism for quantifying the vortex ellipticity by visualizing the two-dimensional ellipse as a projection of a virtual construct, a circular vortex with an axis of symmetry described by tilt angles $\xi$ and $\theta$ as shown in Fig. \ref{zprime_projection}. With this formalism, vortex ellipticity can be determined without knowing $\psi_{bg}$; instead, the background amplitude will be calculated once the angles of the vortex are known. The task of quantifying such tilt angles for an elliptical vortex within an arbitrary paraxial field is taken up first.

It simplifies the ensuing analysis to treat the complex paraxial field $\psi = \psi_R + i \psi_I$ as a two-dimensional vector field $\vecu(\vecr,z)$ mapped onto the transverse plane, where $\vecr$ is the transverse position vector. For instance, the field of a circular linear-core vortex at the origin, $\psi_{\scriptscriptstyle v}^{circ} = x +i y$, can be represented as $\vecu_{\scriptscriptstyle v}^{\scriptscriptstyle circ}(\vecr) = \vecr$. This allows us to describe vortex dynamics in terms of the locally homogeneous deformations of a vector-valued continuum. The paraxial equation is now
\begin{equation}\label{parax_vec}
	\pauli \partial_z \vec u = -\frac{1}{2} \nabla^2_\perp \vec u + {\cal V}\vec u,
\end{equation}
where a Pauli-like operator $\pauli$ is the natural counterpart to $i$: 
\begin{equation}\label{pauli}
	[\sigma_0]=\begin{pmatrix}
		0 & -1 \\
		1 & 0  
	\end{pmatrix}.
\end{equation}

To produce an isolated elliptical vortex in this setting, the circular symmetric reference field $\vecu_{\scriptscriptstyle v}^{\scriptscriptstyle circ}(\vecr)$ is deformed using a homogeneous, in-plane deformation $\FLC$ to give the new vector field, 
    \begin{equation}
        \vecu_{\scriptscriptstyle v}(\vecr) :=\vecu_{\scriptscriptstyle v}^{\scriptscriptstyle circ}(\FLCinv\vecr) \equiv \FLCinv \vecu_{\scriptscriptstyle v}^{\scriptscriptstyle circ}(\vecr) \equiv \FLCinv \vecr .
        \label{eq:utoFinvr}
    \end{equation}
With circular vortices viewed as the reference configuration of a two-dimensional (2D) continuum, elliptical vortices are obtained through a homogeneous deformation by $\FLCinv$. The vector field $\vecu_{\scriptscriptstyle v}$ can thus be more compactly visualized as a deformed continuum instead of assigning a distinct vector to each position.

The polar decomposition theorem allows such 2D deformations to be decomposed into axial stretches or compressions along with a rigid rotation. In particular, $\FLC = \RLC\ULC$, where $\ULC$ is a symmetric deformation (axial stretch or compression) and $\RLC$ is a rigid rotation. This is equivalent to $\FLC = \VLC\RLC$ where now the same rotation is applied first and is followed by a different symmetric deformation $\VLC$ that can be written as $\VLC=\RLC \ULC \RLC^{-1}$. In our consideration of vortices, it is sufficient to consider a single uniaxial stretch, allowing deformations to be completely described by two angles in the eigenbasis of $\ULC$:
\begin{equation}\label{UR}
[U_{\scriptscriptstyle v}] =\begin{pmatrix}
1 &  0 \\
0& \sec\theta  
\end{pmatrix} ,
\quad 
[R_{\scriptscriptstyle v}] =\begin{pmatrix}
\cos\xi &  -\sin\xi \\
\sin \xi & \cos\xi  
\end{pmatrix} .
\end{equation}
The total vortex deformation, $\FLC$, is then represented by the matrix 
\begin{equation}\label{F0}
[F_{\scriptscriptstyle v}] =[R_{\scriptscriptstyle v}] [U_{\scriptscriptstyle v}] =  \begin{pmatrix}
\cos\xi &  -\sin\xi \sec\theta \\
\sin\xi& \cos \xi \sec\theta 
\end{pmatrix} .
\end{equation}

A useful way to visualize a vortex deformed according to Eq. (\ref{F0}) is to view $\xi$ and $\theta$ as describing the orientation of a virtual construct, a circular vortex with a symmetry axis that is tilted with respect to the direction of beam propagation.  This is shown in Fig. \ref{zprime_projection}, where a primed coordinate system is introduced with the symmetry axis of the construct as $z'$. Projection onto the transverse plane is carried out along this axis. The primed frame is obtained by rotating the original coordinate frame by $\xi$ about the $z$ axis followed by a rotation of the frame by $\theta$ about the new $x$ axis. The Euler matrix associated with these two rotations is
\begin{equation}\label{Rtot}
[R_{\rm 3D}] =\begin{pmatrix}
\cos\xi &  -\cos\theta\sin\xi & \sin\theta\sin\xi\\
\sin\xi & \cos\theta\cos\xi & -\cos\xi\sin\theta \\
0 & \sin\theta & \cos\theta  
\end{pmatrix}.
\end{equation}
A comparison of Eqs. (\ref{F0}) and (\ref{Rtot}) shows that the inverse transpose of planar deformation $\FLCinvtrans$ is just the ($x,y$) projection of a three-dimensional rotation ${\bf R}_{\rm 3D}$. A vortex in the primed plane has an elliptical projection in the original transverse plane, and vortices are completely characterized by their in-plane position and two angles of orientation $\{{\bf r}_\perp, {\bf R}_{\rm 3D} \}$. This \emph{virtual tilted vortex} perspective simplifies analyses and offers intuition into the understanding and control of vortex dynamics. 

%FIGURE # 1
%
%
% Vortex_Tilt_and_Velocity_122320.nb

The continuum analysis of tilted vortices can be extended to quantify vortex orientation associated with an arbitrary paraxial field. Take the gradient of $\psi = \pbg \pLC$ and evaluate the result at the vortex center. Since $\pLC = 0$ there, this gives $\gs\psi = \psi_{bg} \gs\pLC$, which is equivalent to
\begin{equation}\label{key}
\nabla_\perp \vecu = \rho_{bg}\Phibg^T \nabla_\perp \uLC .
\end{equation}
Here the background field is expressed in terms of its amplitude and phase, $\psi_{bg} = \rho_{bg}e^{i \varphi_{bg}}$, and the operator $\Phibg$ generates vector rotations by $\varphi_{bg}$,
\begin{equation}\label{PhiBGmat}
	[\Phi_{bg}]=
	\begin{pmatrix}
		\cos\varphi_{bg} & \sin\varphi_{bg} \\
		-\sin\varphi_{bg} & \cos\varphi_{bg}  
	\end{pmatrix} .
\end{equation}
As can be seen from Eq. (\ref{eq:utoFinvr}), the gradient field, $\nabla_\perp \uLC$ is just the inverse deformation gradient, $\FLCinv$. Similarly, $\F^{-1} :=\nabla_\perp \vecu$ is the inverse deformation gradient of the total paraxial field.  With the former evaluated at the vortex center, we can write
% The gradient fields, $\nabla_\perp \vecu$ and $\nabla_\perp \uLC$, are just inverse deformation gradients, $\F^{-1}$ and $\FLCinv$, with the former evaluated at the vortex center to give
\begin{equation}\label{Frelation}
\F = \frac{1}{\rho_{bg}}\FLC\Phibg .
\end{equation}
In principle, it would be possible to use Eq. (\ref{Frelation}) to find both $\xi$ and $\theta$, but this would be a difficult undertaking since the background amplitude and phase are obtained by removing a vortex with the very tilt angles sought. A different strategy is therefore taken. Recalling the polar decomposition $\F = \V\R$, with $\Vsq =\F \F^T$, the analogous symmetric deformation for the vortex field is $\VLC^2=\FLC \FLC^T$. Performing the calculation of $\F \F^T$ using Eq. (\ref{Frelation}) and noting that both $\RLC$ and $\Phibg$ are unitary yields
\begin{equation}\label{Vsqmateq}
\Vsq = \rho_{bg}^{-2}\VLCsq .
\end{equation}
The orientation of an arbitrary paraxial vortex can therefore be described in terms of the orientation of a circular linear-core vortex along with the local background amplitude. While it was convenient to describe vortex stretching in terms of $\ULC$, there is an immediate benefit in calculating $\Vsq$ due to its independence from the background phase.

The eigenvalues of $\VLCsq$ are $\rho_{bg}^2$ and $\rho_{bg}^2 \sec^2\theta$ and the associated eigenvectors are $\{\cot\xi,1\}$ and $\{-\tan\xi,1\}$. The ratio of the eigenvalues and the eigenvectors is then only in terms of the unknown tilt angles, removing the need for the background amplitude to be known. However, the eigensystem of the left-hand side is completely determined by the paraxial field $\psi$: One can calculate $\Vsq$ from the total paraxial field by using the relationships between $\Vsq$, $\F$ and $\nabla_\perp \psi$, as in Appendix \ref{Appendix:tiltmeas}. Denote the known eigenvectors of $\Vsq$ as ${\boldsymbol\nu}_j$ with eigenvalues $\lambda_j$, $j = 1,2$. Then the azimuthal rotation $\xi$ and polar dip $\theta$ are given by
\begin{equation}\label{tilt_final}
\theta= \cos^{-1}\left(\pm\sqrt{\frac{\lambda_{1}}{\lambda_{2}}}\right), \quad
\xi = \cot^{-1}\left(\frac{\nu_{1x}}{\nu_{1y}}\right) + n \pi,
\end{equation}
where $\lambda_1 < \lambda_2$ and $n\in\mathbb Z$. The signs can be determined from the sign of the components of Eq. (\ref{Frelation}).

We have found that the paraxial field local to any vortex can be used to quantify its orientation. To summarize the process, it is possible to find the tilt of the circular construct to describe vortex ellipticity by: (i) using derivatives of the total paraxial field to construct $\Vsq$ (see Appendix \ref{Appendix:tiltmeas}), (ii) finding the eigenvalues and eigenvectors of $\Vsq$, and (iii) using Eq. (\ref{tilt_final}) to solve for $\xi$ and $\theta$. After the tilt of the circular-vortex construct is obtained, the background field can be obtained via $\psi_{bg}=\psi/\psi_v$, which will later be used in Sec. \ref{sec:velocity}. 

We now turn to realizing and measuring such vortices in a laboratory setting.

%%%%%%%%%%%%%%%%%%%%%%%%%%%%%%%%%%%
\subsection{Experimental Generation of Tilted Optical Vortices}
%%%%%%%%%%%%%%%%%%%%%%%%%%%%%%%%%%%

\begin{figure}[t]
\begin{center}
\includegraphics[width=\linewidth]{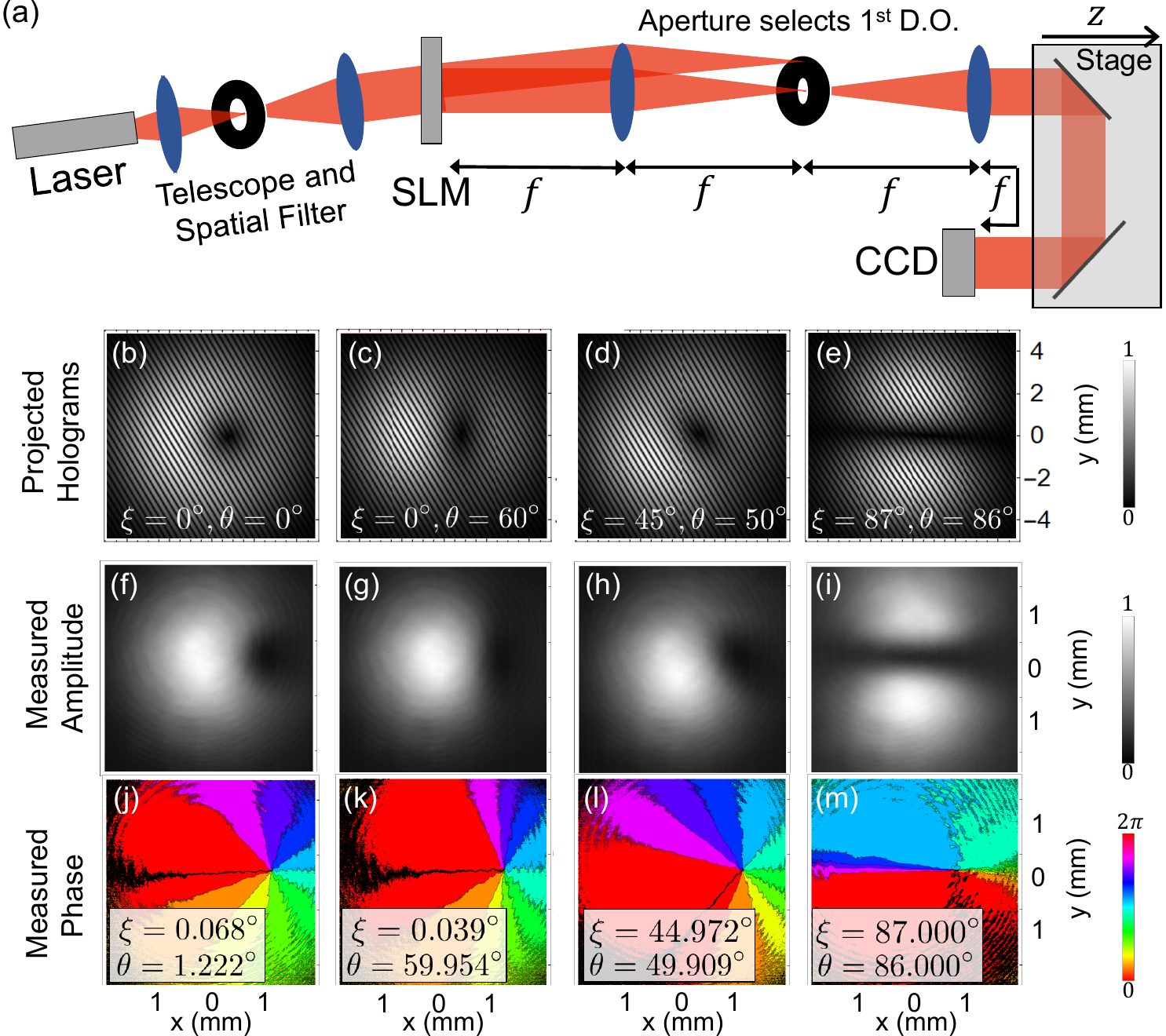}
\end{center}
\caption{ \textit{Experimental generation and measurement of a tilted optical vortex.}  (a) Schematic of the experimental setup. Here DO denotes diffracted order. (b)-(e) Projected holograms for vortices with $x_0=0.8$ mm and four sets of encoded tilt angles with corresponding measured amplitudes (f)-(i), and phases (j)-(m). A 2D amplitude fit of (f) at $z=0$ mm using Eq. (\ref{Single}) gives a measurement of $w_0=1.1$ mm. (j)-(m) show measured tilt angles (insets) at $z=0$ mm calculated by fitting the phase data to Eq. (\ref{Single}) with fitting parameters $x_0$, $y_0$, $\xi$ and $\theta$.} \label{fig:single_vortex_and_setup} 
\end{figure}

A laser beam is an optimal 2+1D medium  in which vortices with a specified tilt can be embedded. In our experiments, laser light is transmitted through a spatial light modulator (SLM) that displays a hologram. The first diffracted order is collected through a $4f$ imaging system with an aperture, and a translation stage controls the propagation distance before the beam is measured by a CCD camera. The complete experimental set up is shown in Fig. \ref{fig:single_vortex_and_setup}(a).

As an example, the field of a single tilted vortex placed at location $r_0$ within a host Gaussian beam is given by 
\begin{equation}
    \psi_0(x,y) = \pbg(x,y,0) \vecu_{\scriptscriptstyle v}(\vecr-\vecr_0),
    \label{eq:singlevortexexperimental}
\end{equation}
where $\pbg(x,y)$ is the same as in Eq. (\ref{LC}). Utilizing the relationship $\vecu_{\scriptscriptstyle v}(\vecr-\vecr_0)= \FLC^{-1}(\vecr-\vecr_0)$, the full field of a tilted vortex in a Gaussian beam at $z=0$ is given by
\begin{align*} \label{Single}
    \psi_0(x&,y) =\sqrt{\frac{1}{\pi}} \biggl\{ [(x - x_0) + i (y - y_0) \cos{\theta}]\cos{\xi}\\
    &+[(y - y_0) - i (x - x_0) \cos{\theta}] \sin{\xi} \biggl\} e^{-(x^2+y^2)/2}, \numberthis
\end{align*}
where $(x_0,y_0)$ is the nondimensional position of the vortex and the characteristic transverse length is $w_0/\sqrt{2}$. A dimensional expression is used for the experiment (see Appendix B). The field is superimposed with a tilted plane wave to create a hologram which will shape the beam at $z=0$. Example holograms demonstrating the control of vortex tilt are shown in Figs. \ref{fig:single_vortex_and_setup}(b)-(e). 

%
%FIGURE # 2
%
%%%% 
%D:\Publication Info and Figures_Tilt Paper\Background Field and individual contributions to velocities figures.nb

To confirm the fidelity of our tilted vortices, the stage is moved to the imaging plane where both the amplitude and phase are measured with collinear phase-shifting digital holography~\cite{andersen2019characterizing}. In addition to the holograms shown in in Fig. \ref{fig:single_vortex_and_setup}, the phase measurement of each vortex requires four additional holograms with a phase stepped reference mode (see Appendix B).

Experimental measurements of the field amplitude and phase at the imaging plane are shown in Fig. \ref{fig:single_vortex_and_setup}(f)-(m). The measured tilt angles show excellent agreement with the programmed tilt angles. The measured mode size is smaller and the vortex is closer to the edge of the beam than in the hologram because the output of the hologram is the product of the Gaussian within the hologram and the incident Gaussian beam. The result is a smaller generated host Gaussian and a larger apparent vortex offset for the measured mode, but vortex tilt and offset from center are unaffected.

We have shown that tilted vortices can successfully be generated and measured in laser beams. The next step is to consider the evolution of such vortices with propagation and develop a complementary characterization of vortex velocity.

%%%%%%%%%%%%%%%%%%%%%%%%%%%%%%%%%%
\section{Vortex Kinetic Equation that Includes Tilt}
\label{sec:velocity}
%%%%%%%%%%%%%%%%%%%%%%%%%%%%%%%%%%

 Consider a general paraxial vector field $\vec u(x,y,z)$ that contains a single moving vortex with position described by ${\vecr}_v(z)$. The vortex is restricted to have unit charge. Define the vortex velocity by ${\vec v} = \partial_z{\vecr}_v$. At the vortex center, $\vecu = \vec 0$. Because the value of the paraxial field at the vortex location is always equal to zero, and therefore a constant, it is guaranteed that $d_z\vecu=0$ for all $z$. The total derivative of $\vecu$ with respect to $z$ can also be expressed as $d_z\vecu= \partial_x \vecu  \hspace{0.04in} \partial_z x+\partial_y \vecu \hspace{0.04in}\partial_z y +\partial_z \vecu$ $=\nabla_\perp \vecu \hspace{0.04in}\partial_z \vecr_v +\partial_z \vecu$. Using the relationships just mentioned, the total derivative is
\begin{equation}\label{v0}
d_z \vec u = \F^{-1}\vec v + \partial_z \vec u = \vec 0,
\end{equation}
with the transverse position evaluated at ${\vecr}_v(z)$ for all fields and where $\F^{-1} \equiv\nabla_\perp \vecu$. This implies that the vortex velocity is
\begin{equation}\label{v1}
{\vec v} = -\F \partial_z \vec u .
\end{equation} 
We rewrite this using Eqs. (\ref{parax_vec}) and (\ref{Frelation}) to obtain
\begin{equation}\label{v2}
{\vec v} = \frac{1}{2 \rho_{bg}} \FLC \Phibg\pauli^{-1} \nabla^2_\perp \vecu.
\end{equation}
Because this is evaluated at the vortex center, the potential term ${\cal V} \vec u$ drops out. Such terms can affect the vortex dynamics of course, but only through their influence on the background field.

This expression for vortex velocity can be simplified by noting that a vortex located at ${\vecr}_v$ is described by $\vecu_{\scriptscriptstyle v} = \FLCinv({\vecr}-{\vecr}_v)$ so $\vecu = \rho_{bg}\Phibg^T\FLCinv({\vecr}-{\vecr}_v)$. The Laplacian term of Eq. (\ref{v2}), evaluated at the vortex center, can therefore be written as
 \begin{equation}\label{lap}
 \nabla^2_\perp \vecu = 2 \nabla_\perp \cdot (\rho_{bg} \Phibg^{-1} \FLC^{-1} ) .
 \end{equation}
Substitution of this result into Eq. (\ref{v2}), as well as a laborious reduction, gives
\begin{equation}\label{v_final}
{\vec v} = \nabla_\perp \varphi_{bg} - \frac{\VLCsq}{\JLC}\pauli\nabla_\perp \mathrm{ln}\rho_{bg},
\end{equation}
%Rings_Fields_2D_Projections_042520.nb
where $\JLC = {\rm Det}(\FLC)$.

For incompressible materials, this equation is consistent with a hydrodynamic interpretation of the paraxial equation~\cite{Madelung1927}, i.e., it is just the phase gradient, implying that vortices move with the underlying fluid. However, light exhibits compressibility, so vortex motion is also influenced by local gradients in fluid density, the square of the paraxial field magnitude. In the absence of tilt, the equation above reduces to a form previously obtained in the context of the Gross-Pitaevskii equation for quantum fluids~\cite{Nilsen2006, Groszek2018} in which such density gradients are accounted for. 

Equation (\ref{v_final}) goes a step further though, revealing the coupling between the vortex tilt and local gradients in the hydrodynamic density. This coupling is strongest when the local density gradient points in the direction of the eigenvector of $\VLCsq$ with unity eigenvalue. Tilt decreases the vortex speed in this direction by a factor of $\cos\theta$ while increasing the speed in the direction of tilt by a factor of $\sec \theta$. This implies that, as tilt approaches $90^\circ$, the vortex speed in the direction of tilt will approach infinity, provided the untilted velocity is finite in this direction, while that in an orthogonal direction approaches zero.

In contrast, there is no such coupling between the vortex phase gradient and its tilt. Returning to the hydrodynamic perspective, phase gradients are the fluid velocity that may sweep vortices along like a keeled boat drifting downriver without regard for its heading. An amplitude gradient, though, corresponds to a fluid pressure differential, which causes a vortex to move relative to the underlying fluid. Now the orientation of the keel is relevant as the relative motion between the vessel and fluid could cause the boat to cut a path that is not necessarily in the direction of lower pressure.

The combination of Eqs. (\ref{tilt_final}) and (\ref{v_final}) comprises a coupled system that can be solved to predict the evolution of vortex position and tilt. Significantly, they hold for both linear and nonlinear optical media. If axial location $z$ is replaced by time $t$ the system can be applied directly to predict the dynamics of vortices in quantum fluids within the Gross-Pitaevskii ansatz~\cite{Pethic}.

We next turn to linear optical experiments to verify the predictive power of this model by looking carefully at two specific test cases.

%%%%%%%%%%%%%%%%%%%%%%%%%%%%%%%%%%%%%%%%%%%%%%%%%%
\section{Dynamics and Tilt Evolution in Two Cases}
%%%%%%%%%%%%%%%%%%%%%%%%%%%%%%%%%%%%%%%%%%%%%%%%%%

%%%%%%%%%%%%%%%%%%%%%%%%
\subsection{Case 1: Single Off-Center Tilted Vortex in a Gaussian Beam}
\label{sec:SingleVortex}
%%%%%%%%%%%%%%%%%%%%%%%%

In the absence of any tilt, a vortex placed on the side of a Gaussian beam will initially move perpendicular to its radial line~\cite{Rozas1999}, but the full trajectory has not been identified and the influence of tilt is yet to be elucidated. Towards this end, consider the initial state of a paraxial field constructed as the product of a vortex, placed on the $x$ axis at $x_0$ with initial tilt $\{\xi_i, \theta_i\}$, and a Gaussian beam of waist $w_0$ and wave number $k$. Position is nondimensionalized with $w_0/\sqrt{2}$ (inplane) and Rayleigh length (propagation axis). A convolution with the paraxial Green's function~\cite{FourierFresnelPaper} is used to derive the field as a function of position along the $z$ axis where
\begin{align*}\label{paraxial_SVG}
& \psi(x,y,z) =[\cos\xi_i(-x+x_0+i x_0 z - i y \cos\theta_i)+ ... \\
&\hspace{0.35in}- y \sin\xi_i+ i x \cos\theta_i\sin\xi_i + x_0(-i+z)\cos\theta_i\sin\xi_i]\\
&\hspace{0.5in} \times \frac{1}{(-i + z)^2}\sqrt{\frac{2}{\pi}}e^{i(x^2+y^2)/((2(-i+z))}.\numberthis
\end{align*}
% Single_Vortex_on_Gaussian\Mathematica\Tilted_Single_Vortex_Gaussian_052020.nb

The coupled Eq. (\ref{tilt_final}) and (\ref{v_final}) can then be solved to find two key results. The first is that vortex tilt does not evolve from its initial value, as confirmed by calculating both $\xi$ and $\theta$ as a function of $z$. Second, the vortex position is given by:
\begin{eqnarray}\label{position_SVG}
x_v &=& x_0 - x_0 z \cos\xi_i\sin\theta_i\sin\xi_i\tan\theta_i \nonumber\\
y_v &=& x_0 z \cos\theta_i\big(\cos^2\xi_i \sec^2\theta_i + \sin^2\xi_i\bigr) .
\end{eqnarray}
% Single_Vortex_on_Gaussian\Mathematica\Tilted_Single_Vortex_Gaussian_052020.nb
%
Here Eq. (\ref{v_final}) has been integrated to construct this vortex trajectory. The result is consistent with the trajectory calculated from intersections of real and imaginary zeros of the propagated mode, confirming the validity of Eq. (\ref{tilt_final}) in this optical setting. 
%
%FIGURE # 3
%
\begin{figure}[t]
	\begin{center}
		\includegraphics[width=\linewidth]{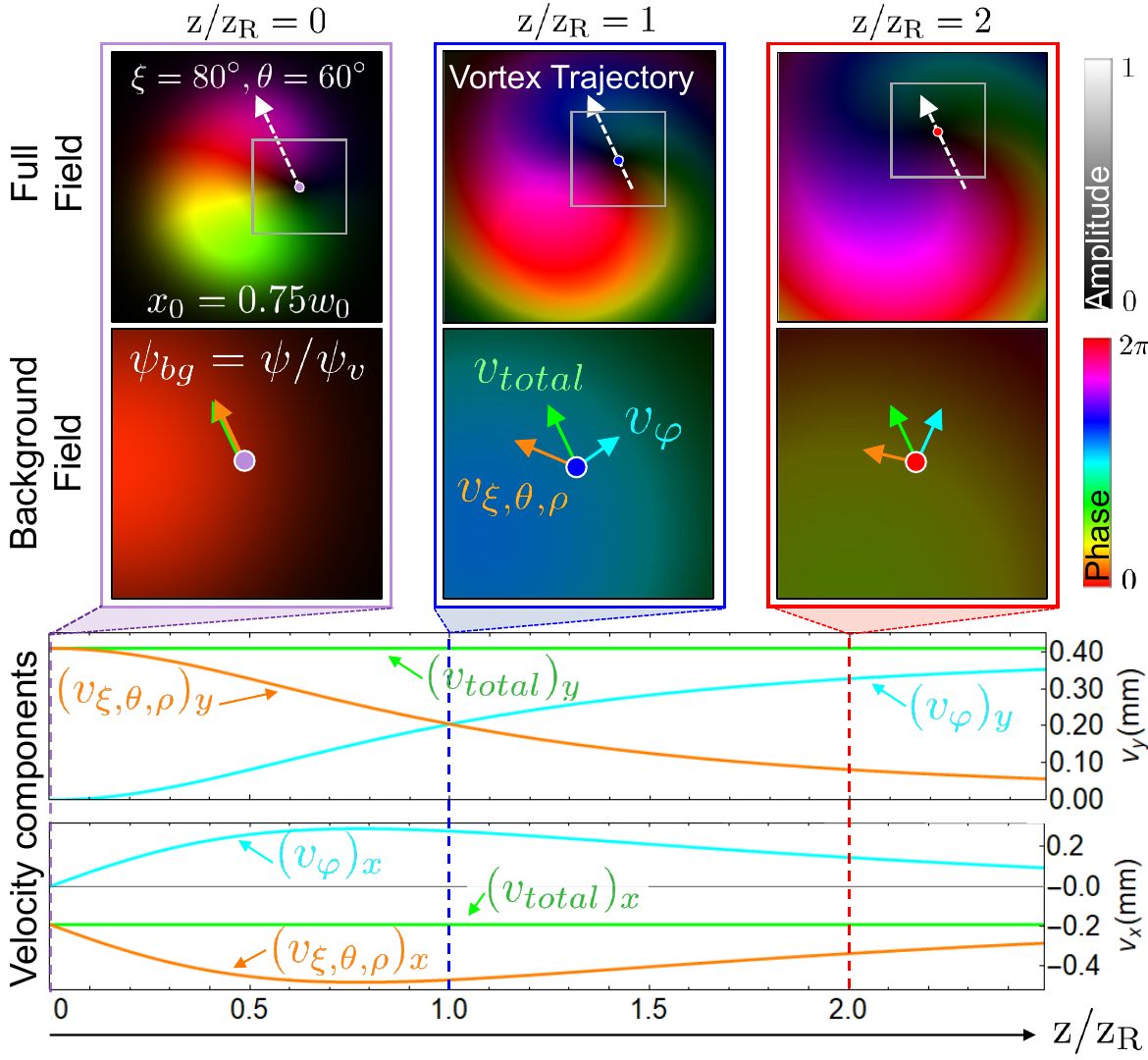}
\end{center}
\caption{ \textit{Motion of a single vortex driven by tilt and background field gradients.} Shown on top are the mode snapshots of an off-center, tilted vortex calculated from Eq. (\ref{paraxial_SVG}) with its location (dots) and predicted trajectory (white arrow) for beam parameters $\lambda=633$ nm, $w_0=1$ mm, $x_0=0.75w_0$, $\xi_i=80^\circ$ and $\theta_i=60^\circ$. Brightness is amplitude and color shading represents phase. The middle shows a cropped window of the background field evolution, along with background gradient vectors to visualize each contribution to vortex velocity: cyan (farthest right) arrows show velocity from the first term of Eq. (\ref{v_final}), $v_{\varphi}$, while orange (farthest left) arrows shows the second term, $v_{\xi,\theta,\rho}$; green (middle) arrows are the sum of these terms ($v_{total}$). Shown on the bottom is the evolution of individual components and resulting total velocity for both the $x$ and $y$ directions, labeled by the corresponding component. 
\label{fig:singlevortexcomp}
}
\end{figure}
%%% 
%D:\Publication Info and Figures_Tilt Paper\Velocity Components and Background Field_Single, Tilted, Offcenter Vortex_02_07_21.nb

Surprisingly, we see from Eq. (\ref{position_SVG}) that the calculated vortex velocity with propagation is a constant for this single-vortex case. The background field can be determined by using the vortex position from Eq. (\ref{position_SVG}), with $\xi$ and $\theta$ calculated as described in Sec. \ref{sec:tiltformalism}, and dividing out the total elliptical vortex from the paraxial field of Eq. (\ref{paraxial_SVG}). Figure \ref{fig:singlevortexcomp} shows a visual depiction of the evolving optical mode and vortex trajectory for a vortex with tilt angles $\xi_i=80^\circ$ and $\theta_i=60^\circ$, where the velocity is clearly seen to be driven by the phase and tilt plus amplitude gradient terms in Eq. (\ref{v_final}). Although the vortex velocity remains constant, the evolution of the background field changes the relative strength of the background phase and amplitude gradients such that changes in both the $x$ and $y$ components perfectly balance throughout the entire beam propagation, resulting in a constant total velocity. We find that this surprising constant velocity results from the equally-surprising vortex decomposition of the propagating field:  The background field is a simple diverging Gaussian and changes in the amplitude and phase gradients exactly cancel.\
%
%FIGURE # 4
%
\begin{figure}[h!]
	\begin{center}
		\includegraphics[width=0.85\linewidth]{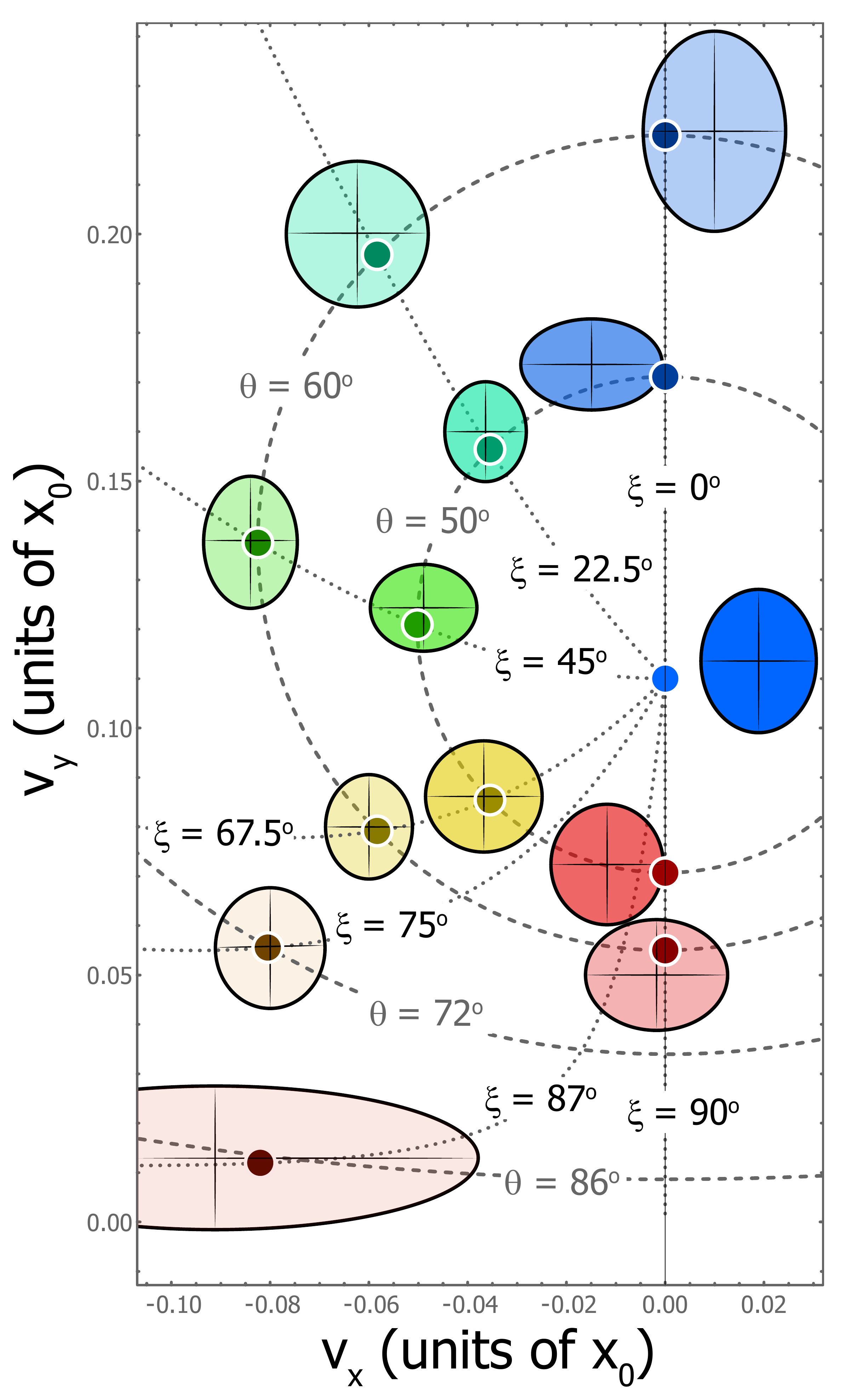}
	\end{center}
	\caption{ \emph{Vortex on shoulder of the Gaussian beam}. Velocity predictions in mm/m for both the $x$ and $y$ directions are shown for a single vortex with various tilts. Small circles are theoretical predictions from Eq. (\ref{position_SVG}) and crosshairs are the measured vortex velocities with elliptical regions bounding the experimental uncertainty. Dashed semicircles show the predicted velocities for a specified polar angle $\theta_i$ as the azimuthal orientation $\xi_i$ is varied. Dotted lines are predictions for a fixed $\xi_i$ and varying $\theta_i$. In the absence of tilt, the vortex moves vertically, but one should note that specific combinations of tilt can also lead to a purely upward velocity. For severe tilt ($\theta_i$ near $90^\circ$), the vortex can be made to move essentially orthogonal to its untilted trajectory.  }
	\label{Single_Vortex_Gaussian_2}
\end{figure}
%
% C:\Users\mlusk\Documents\Research\Topological_Fluids_of_Light\Mathematica\Vortex_Tilt_Helicity\Expt_Single_Tilted_Vortex\Round_3\Expt_Data_Analysis_Round_3_Single_Tilted_Vortex.nb
% C:\Users\mlusk\Documents\Research\Topological_Fluids_of_Light\Investigations\Vortex_Tilt\Single_Vortex_on_Gaussian\Phase-fitting Vortex Measurement\Velocity_Analysis\Single_Vortex_Gaussian_Expt_Theory_Manuscript_Figure_103120.nb
%

The impact that vortex tilt has on vortex motion for an off-center vortex in a Gaussian beam is quantified in Fig. \ref{Single_Vortex_Gaussian_2}. The predicted velocities for 13 sets of tilt angles are shown with small circles. In the figure, polar angles are limited to $\theta_i\in[0,90]^\circ$ because greater values would correspond to a vortex of opposite charge. Negatively charged vortices move in the direction opposite to their positive counterparts. Interestingly, the vortex velocity is an even function of azimuth angle $\xi_i$ despite the initial vortex position being offset from the beam center. Also note that the vortex velocity can never be brought to zero via tilt. These results can be compared directly with experimental measurements using the apparatus and holograms summarized in Fig. \ref{fig:single_vortex_and_setup}.

Thirteen experiments were carried out in which a vortex was initially offset by the same $x_0=0.8$ mm from the center of a $w_0=1.1$ mm Gaussian laser beam for various tilts. To measure the trajectory of a given vortex, mode measurements were made as the translation stage in Fig. \ref{fig:single_vortex_and_setup} was stepped back. The vortices were located by computationally identifying intersections of the real and imaginary zeros of the experimental field~\cite{gbur2016singular}. Measurements of a Laguerre Gaussian beam as in Eq. (\ref{LC}) were also made at each propagation step to remove any extraneous motion due to beam drift. 

\begin{figure}[t]
\includegraphics[width=0.45\textwidth]{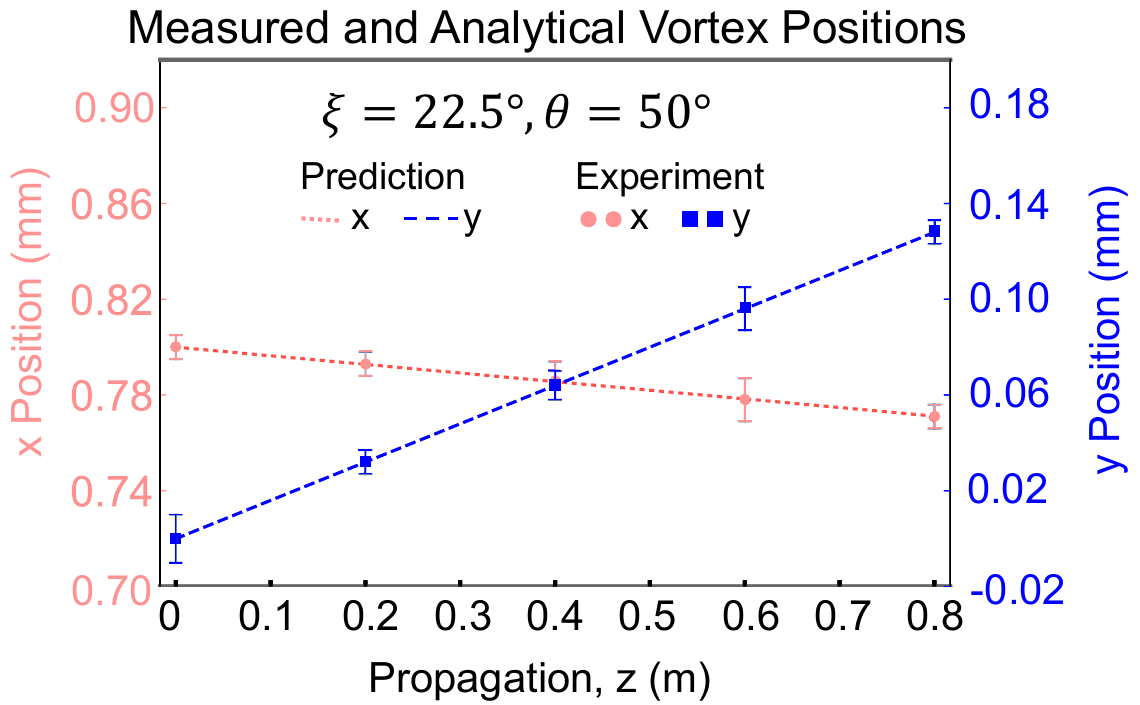}
\caption{ \textit{Experimentally measured and analytically predicted vortex trajectory with $\xi=22.5^\circ$ and $\theta=50^\circ$.} The experimentally measured vortex position as a function of propagation is shown for both the $x$ [in red (light gray)] and $y$ [in blue (black)] directions. Error bars represent one standard deviation of five independent measurements. The dashed lines are the analytical predictions from Eq. (\ref{position_SVG}). Fitted slopes (not shown) calculated using the propagated uncertainties for both the $x$ and $y$ motions are well aligned with the analytical prediction and are used to create a single data ellipse in Fig. \ref{Single_Vortex_Gaussian_2} .
\label{fig:singlevortexpositions}}
\end{figure}

Figure \ref{fig:singlevortexpositions} shows an example of a measured vortex trajectory for a tilted vortex with $\xi=22.5^\circ$ and $\theta=50^\circ$. Dashed lines show the analytical prediction from Eq. (\ref{position_SVG}); the match with data is striking without any fitting parameters. The error bars show one standard deviation across five measurements for each position. Linear fits (not shown) weighted by the unique uncertainties~\cite{taylorerror} are calculated for each trajectory, resulting in $x$ and $y$ component velocity and uncertainty measurements.
%
%FIGURE # 5
%
%%%% Experimental Data in Teams
% TFL -> Tilt (1- and 2-Vortex Repository -> Publication Data and Figures

Figure \ref{Single_Vortex_Gaussian_2} also shows the resulting comparison of predictions with experimental velocity measurements based on linear fits of the measured vortex position. The crosshairs show the experimentally measured velocities and the elliptical regions bound the experimental uncertainty measurements.  With the exception of the untilted vortex case, all of the predicted velocities fall within the uncertainties of the measurements. The trajectories along the $v_x=0$ axis with $\xi_i=0^\circ$ or $90^\circ $ have the largest deviations from the predicted values, but even in these data sets the measured trajectories match well with the analytical predictions. The discrepancies are attributed to the lack of programmed tilt in the $\xi_i$ angle, allowing for any extraneous sources of vortex tilt from pixelation or overlapping stray light to steer the trajectory of the vortex.
 
Vortex tilt was measured by fitting the phase data described in Fig. \ref{fig:single_vortex_and_setup}. The average of five measurements at each $z$-step confirm that the vortex tilt does remain constant with propagation, as predicted.

%%%%%%%%%%%%%%%%%%%%%%%%
\subsection{Case 2: Vortex Pair in a Gaussian Beam}
%%%%%%%%%%%%%%%%%%%%%%%%

A two-vortex Gaussian beam provides a strong test of the hydrodynamic interpretation of vortex motion: According to Eq. (\ref{v_final}), each vortex should move according to its tilt and the field gradients of both the background Gaussian and the other vortex. This is in contrast with the simpler case of two-vortex motion in an incompressible fluid such as superfluid helium, in which the background amplitude gradients can be neglected because the vortex core size is much smaller than the typical vortex separation. This is why optical vortex pairs move differently from the familiar Magnus effect for vortex pairs in an incompressible fluid: Instead of same-charge pairs circling each other and opposite-charge pairs propelling each other forward, same-charge optical vortices move as if each were a single vortex within a Gaussian beam, moving with straight line trajectories~\cite{Indebetouw1993,Rozas1999} and opposite-charge pairs move at twice the Magnus-predicted speed before heading towards each other and recombining \cite{Indebetouw1993}, although opposite-charge vortex pair dynamics in laser light have not been experimentally observed.

\begin{figure}[t]
	\begin{center}
		\includegraphics[width=\linewidth]{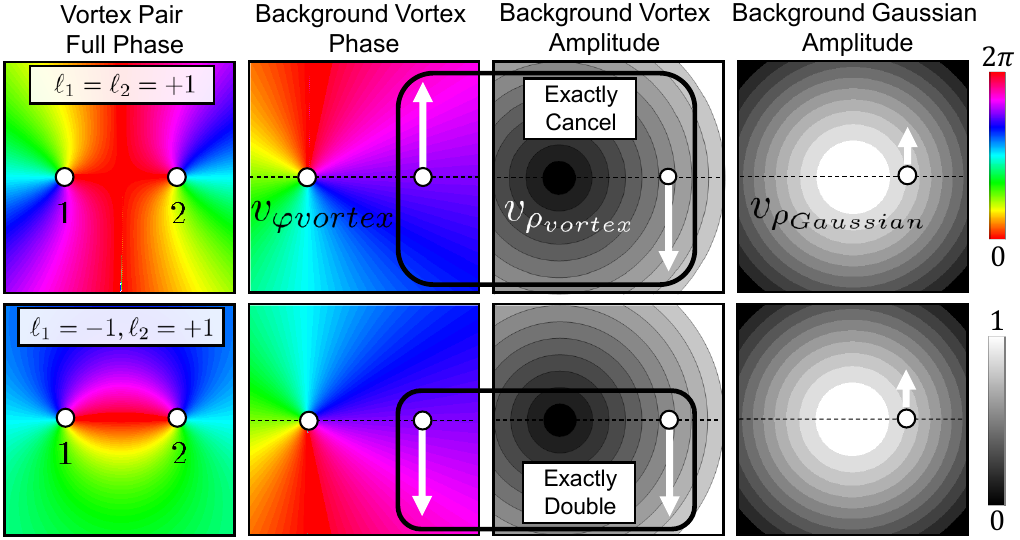}
	\end{center}
	\caption{ \textit{Two vortex motion from background field components at $z=0$.} For a same-charge vortex pair (top row) and opposite-charge pair (bottom row), the background field components for vortex 2 are shown. Columns 2 and 3 show velocity contributions from the background phase and amplitude of vortex 1. The last column shows the background Gaussian amplitude contribution to velocity. }
	\label{fig:twovortexbackground}
\end{figure}

Consider the scenario where a Gaussian beam contains vortex 1 on the left, located at $(-x_0,0)$ and vortex 2 on the right, located at $(+x_0,0)$ at $z=0$, with charge restricted to $|\ell|=1$. If these vortices are initially untilted, the velocity of the right vortex as calculated from Eq. (\ref{v_final}) can be written as:
\begin{equation}
    \vec v_{2,z=0, \textrm{untilted}} = \frac{ \lambda}{2 \pi} \left(  \frac{\ell_1}{2 x_0} -  \frac{\ell_2}{2 x_0}+ \frac{2 \ell_2 x_0}{w^2_0}\right) \hat{y},
    \label{eq:twovortexvelocity}
\end{equation}
where  $\ell_1$ and $\ell_2$ describe the topological charge associated with the orientation of the each vortex: $\theta_v = 0^\circ \rightarrow \ell_v = 1$ and $\theta_v = 180^\circ \rightarrow \ell_v = -1$. The three terms within the parentheses of Eq. (\ref{eq:twovortexvelocity}) describe the phase gradient from vortex 1, the amplitude gradient of vortex 1 (with sign determined by the charge of vortex 2), and the amplitude gradient of the Gaussian beam, respectively. This is visually summarized in Fig. \ref{fig:twovortexbackground}.
%
%FIGURE # 6
%

For the initially circular, like-charge vortex pair, pairwise cancellation of field components implies that each vortex moves in the beam as if the other vortex is not there, vortex 1 moving in along a straight downward line while vortex 2 moves straight upward just as would be expected in the single-vortex case and as is the case for two idealized point vortices in an incompressible fluid~\cite{milne1996theoretical}. This may be why previous authors concluded that like-charge linear core optical vortices exhibit no effective interaction~\cite{Rozas1997}. However, for an initially circular pair with opposite charges, the field gradients exactly add so that each vortex initially moves with twice the Magnus-predicted velocity. This characterization of the initial vortex motion from a hydrodynamic interpretation, including the effect of amplitude gradients and tilt as identified in Eq. (\ref{v_final}), explains a significant difference between two-vortex motion in incompressible fluids and optics.

To verify the model beyond initial conditions, the trajectories and tilt evolution are analyzed next for the unlike charge pair.

\subsubsection{Illustration of Tilted Vortex Evolution and Hydrodynamics}

Consider the nucleation and subsequent annihilation of an oppositely-charged vortex pair. We initialize the system such that at the midpoint $z=0$ the vortices are untilted with charges of $\pm 1$ and are symmetrically separated by $2 x_0$ on a Gaussian beam for consistency with the same $z=0$ field of previous studies~\cite{Indebetouw1993}. The evolving paraxial field can be obtained via convolution with the paraxial Green's function. The dynamics are consistent with the result for $z>0$ previously obtained using Hankel transforms~\cite{Indebetouw1993}.

\begin{figure}[htbp]
	\begin{center}
		\includegraphics[width=\linewidth]{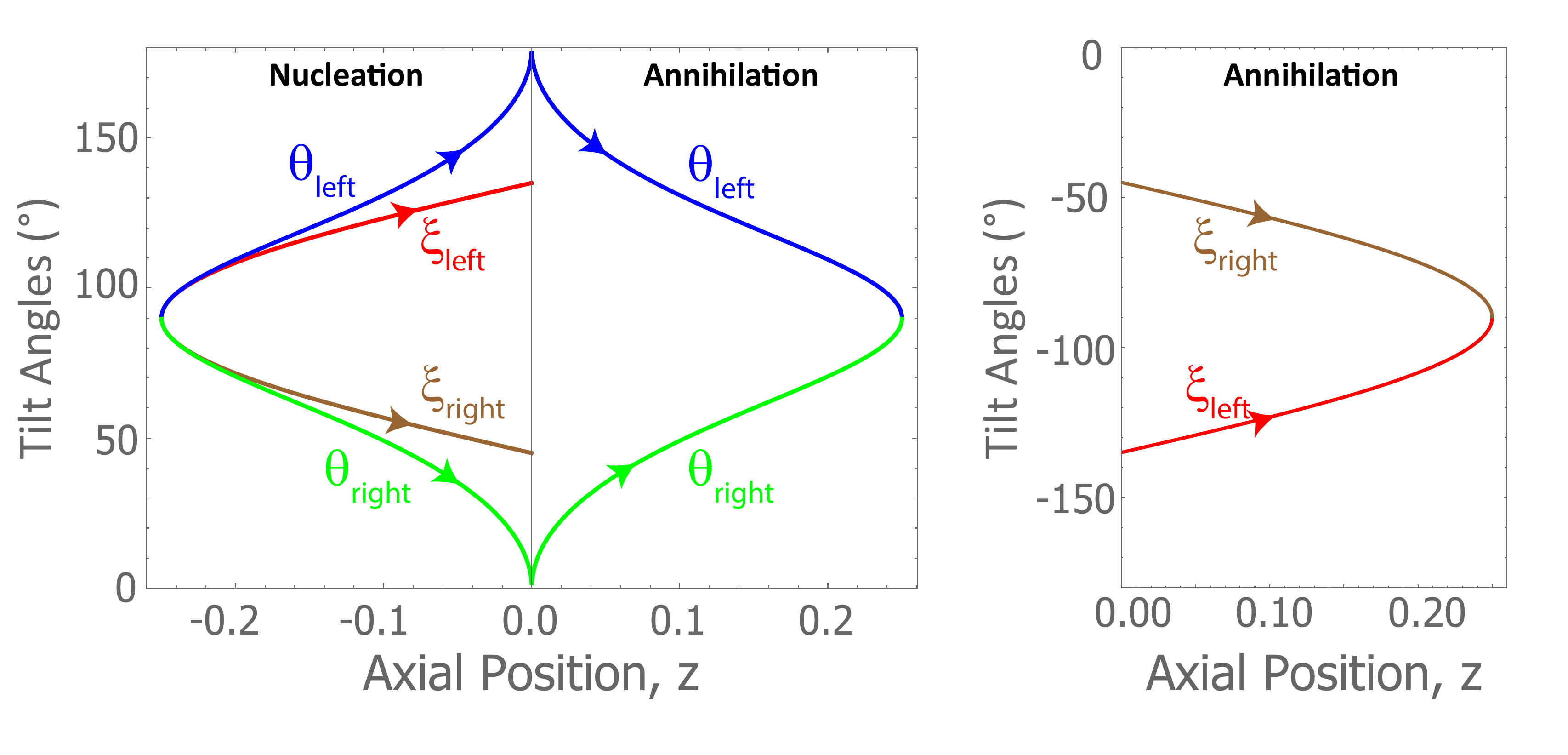}
	\end{center}
	\caption{ \emph{Evolution of tilt for oppositely charged vortex pair as described by Eq. (\ref{tilt_Indeb})}. For the left vortex in the pair, located at $x=-x_0$ at $z=0$, the evolutions of  $\xi_{\rm left}$ and $\theta_{\rm left}$  are shown. Similarly, for the right vortex located at $x=+x_0$ at $z=0$, $\xi_{\rm right}$ and $\theta_{\rm right}$ are also plotted. Each curve is labeled by the corresponding angle.}
	\label{Indeb_1}
\end{figure}

The solution form is particularly simple when considered in the limit of an arbitrarily large beam waist. This leaves only two characteristic lengths, the separation between the vortices $2 x_0$ and the wavelength of the light. Position can then be nondimensionalized using $2 x_0$ (in plane) and $2 k x_0^2$ (propagation axis). The resulting paraxial field, in the limit of a large beam waist wherein the background field becomes uniform, is given by
\begin{equation}\label{paraxial_Indeb} 
\psi_{\scriptscriptstyle v}(x,y,z) = -\frac{1}{4} + x^2 + y^2 + i (y + 2 z)
\end{equation}
The coupled equations (\ref{tilt_final}) and (\ref{v_final}) can then be solved to find, for instance, that for $z\ge 0$ the evolving vortex tilt of the right vortex is given by
\begin{eqnarray}\label{tilt_Indeb}
	\xi(z) &=& \tan^{-1} \frac{-1 -4 z}{\sqrt{1-16 z^2}} \nonumber \\
	\theta(z) &=& \cos^{-1}\sqrt{-1 + \frac{2}{1 + 4 z}} 
\end{eqnarray}
shown in Fig. \ref{Indeb_1}, with an associated trajectory described by
\begin{equation}\label{position_Indeb}
x_v = \frac{1}{2}\sqrt{1-16 z^2}, \quad y_v= -2 z .
\end{equation}
The vortex pair nucleates at $z = -\frac{1}{4}$, annihilates at $z = +\frac{1}{4}$, and traces out a circular trajectory in the ($x,y$) plane.

%
%FIGURE  #7
%

% Momentum_Flux_Indebetouw_052020.nb

\begin{figure}[hptb]
	\begin{center}
		\includegraphics[width=\linewidth]{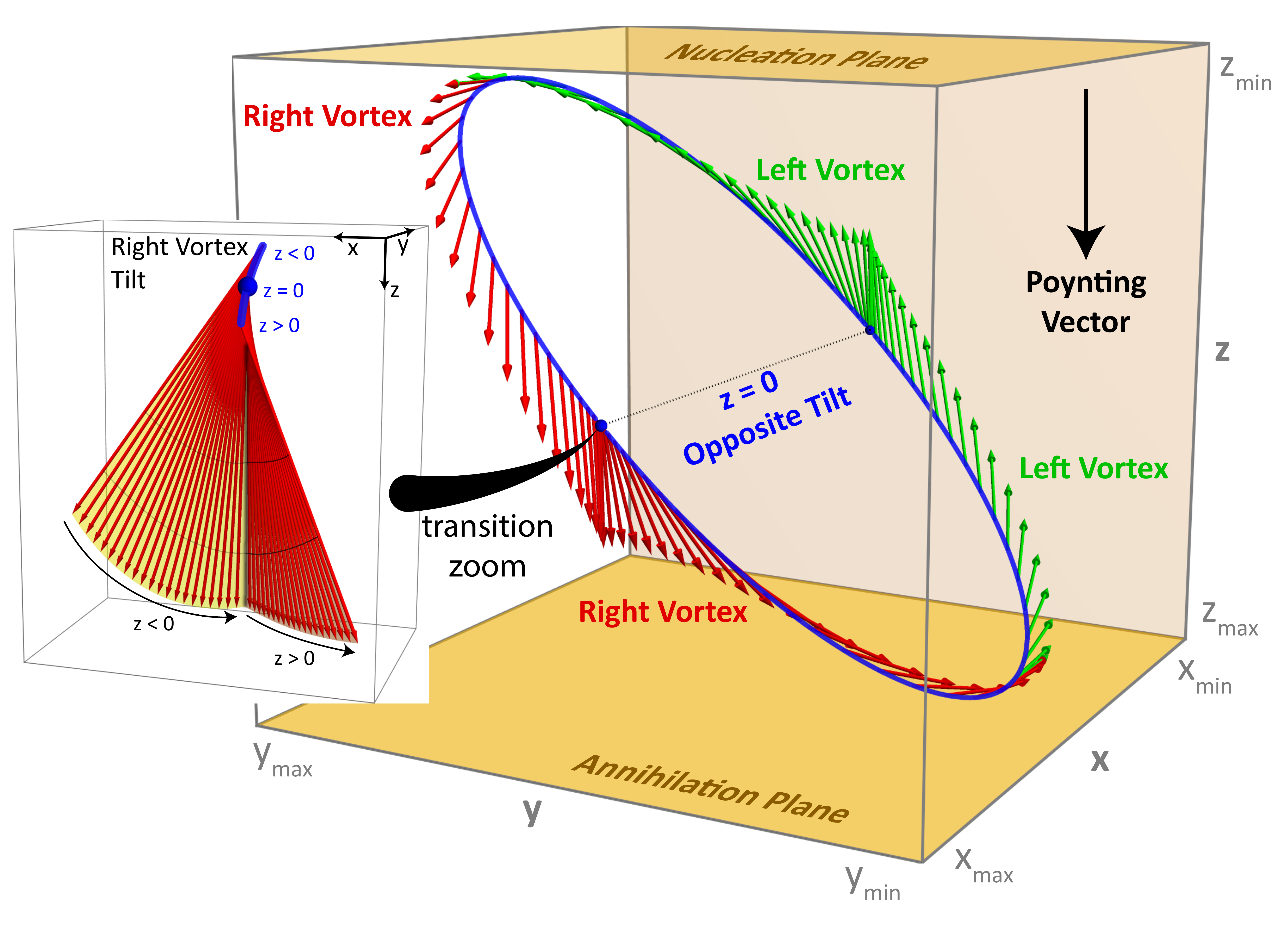}
	\end{center}
	\caption{ \emph{Tilt perspective on nucleation and annihilation.} A pair of identical vortices nucleates at the top left, but subsequently evolves so that the vortices are of opposite charge at the midpoint before becoming once again parallel as they annihilate at the lower right. The envelope of orientations is a M$\mathrm{\ddot{o}}$bius strip. The results shown were obtained by combining Eqs. (\ref{position_Indeb}) and (\ref{tilt_Indeb}) and the tilt vectors shown are of equal magnitude.}
	\label{Indeb_2}
\end{figure}

Knowledge of the evolving orientation allows the background field to be determined via $\pbg = \psi/\pLC$, which can be used to compute the vortex trajectory using Eq. (\ref{v_final}). The result is equal to the axial derivative of Eq. (\ref{position_Indeb}). The combination of position and tilt is used to produce Fig. \ref{Indeb_2}. When phase gradients within a field get sufficiently steep, a vortex pair can nucleate. In particular, when the phase gradient reaches the $\pi$ threshold, two vortices that happen to have the same tilt nucleate~\cite{gorshkov2002topology}, as can be seen in the top left of Fig. \ref{Indeb_2}. The tilts then evolve in opposite directions such that the vortices are of opposite tilt at the mid-point in their life-cycle. They subsequently become re-aligned so that they are once again parallel when annihilation occurs. At both nucleation and annihilation, the polar tilt is $90^\circ$, and the eigenvalue analysis implies that the axial speed must be zero while the lateral speed is infinite. This is consistent with the trajectories plotted. 
%
%FIGURE  #8
%
%
% Momentum_Flux_Indebetouw_052020.nb

 The results of Eqs. (\ref{paraxial_Indeb}) and (\ref{tilt_Indeb}) are the limiting case of an infinite beam waist, but the characteristics of the two-vortex tilt evolution discussed above are also relevant to experiments with finite beams if the initial vortex separation is small compared to the beam waist. A finite beam contributes amplitude and phase gradients that affect the vortex motion, just as in the single-vortex case. Therefore, in our quantitative comparison with experiments below, we include the beam in the Fresnel integration solution of paraxial propagation. The solution is consistent with past work \cite{Indebetouw1993}.

%%%%%%%%%%%%%%%%%
\subsubsection{Experimental Vortex Trajectory and Tilt Evolution}
%%%%%%%%%%%%%%%%

To generate two vortices in a laser beam, a hologram is again created by summing a tilted plane wave with the $z=0$ field with vortex field and background field components given by
\begin{align*}
    &\psi_{\scriptscriptstyle v}(x,y,z) =\biggl[\biggl(x-\frac{1}{2}\biggr)+i y\biggr] \biggl[\biggl(x+\frac{1}{2}\biggr)-i y\biggr]\\
    &\pbg(x,y,z)=e^{-(x^2+y^2)/w_{0,\rm{ND}}^2}.\numberthis
\label{eq:psi0twovortex}
\end{align*}
As usual, all lengths are nondimensionalized using the separation between vortices. A dimensional version of this is used to gethe setup and vortex tracking process are almost identical to those in the single vortex in Sec. \ref{sec:SingleVortex} (see Appendix A). 

\begin{figure}[t]
\begin{center}
  \includegraphics[width=\linewidth]{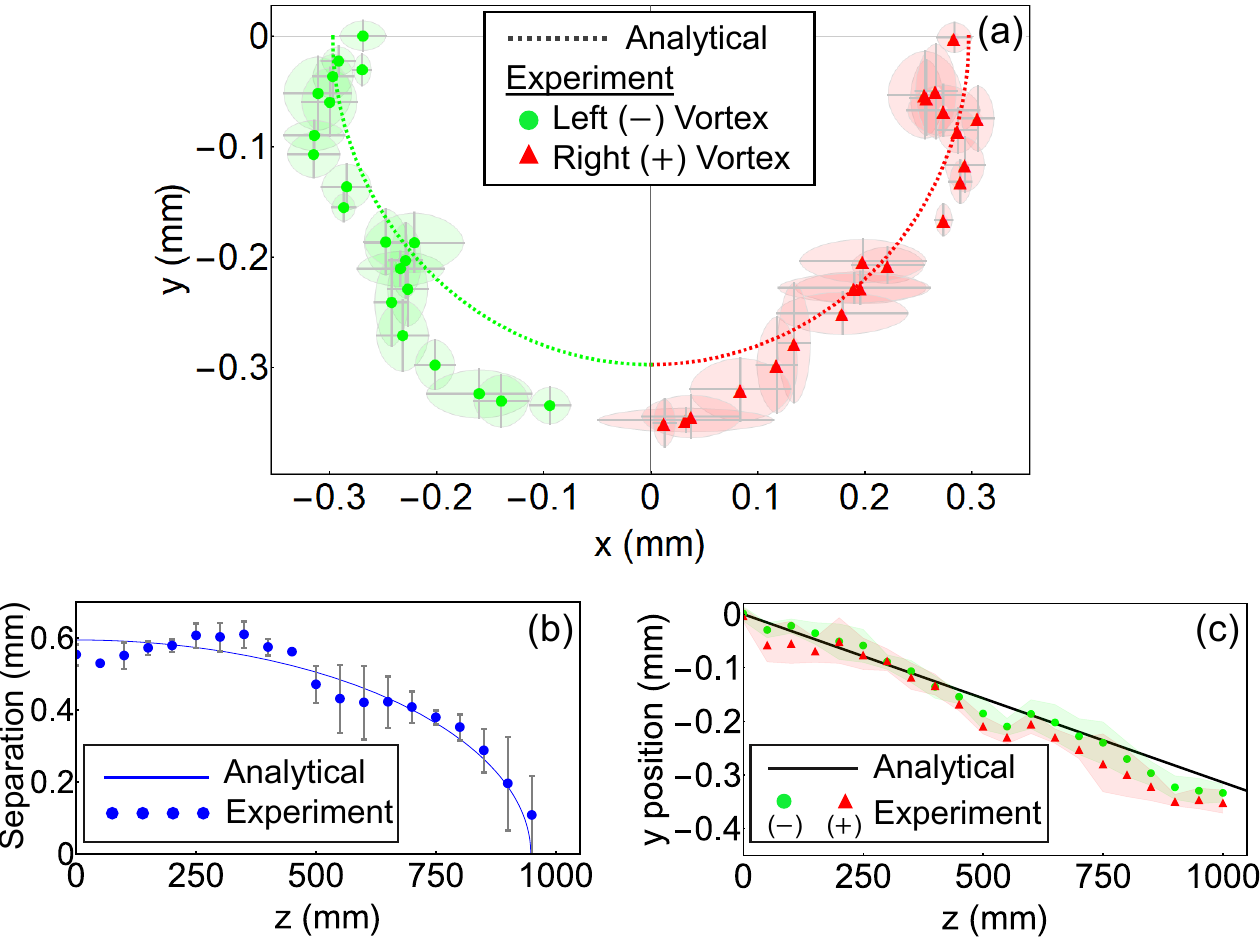}
  \end{center}
  \caption{\emph{Experimental dynamics of opposite charge vortex annihilation.} 
Measured (symbols) and analytical (lines) vortex trajectories for (a) the $x$-$y$ plane measurement for which each green (circle) and red (triangle) symbol pair is a measurement at a specific $z$, (b) the vortex separation along $z$, and (c) the $y$ position as a function of propagation.}
  \label{fig:2vortexexp}
\end{figure}

Measurements of the tracked vortices, plotted in Fig. \ref{fig:2vortexexp}, show the vortex annihilation dynamics.
%
%FIGURE  #9
%
% %%%% Experimental Data in Teams
% % TFL -> Tilt (1- and 2-Vortex Repository -> Publication Data and Figures
%C:\Users\mlusk\Documents\Research\Topological_Fluids_of_Light\Mathematica\Indebetouw_Vortex_Pairs_Expt_Theory\Best_Linear_Core\Tilt_Tracker_Doublet_Annihilation_042520.nb
%
Three representations of the data are shown, each with error bars denoting one standard deviation across five measurements. We observe the anticipated nearly half circular $x$-$y$ trajectory as well as the expected vortex annihilation event at $z \approx 910$ mm. Measurements following the annihilation show no reappearance of the vortices (see Fig. 13 in Appendix C). Using a weighted linear fit [27] on the combined data for $y$ positions of each vortex, the measured $y$ velocity is $v_{y,meas}=-0.364\pm0.009$ mm/m as compared to the predicted $v_{y,model}=-0.314$ mm/m. The slight disagreement can be attributed to systematic error from experimental sources such as beam collimation and roundness, and pixelation, pixel phase error, and finite bit depth on the SLM.

%
%FIGURE #10
%
\begin{figure}[b]
	\begin{center}
		\includegraphics[width=\linewidth]{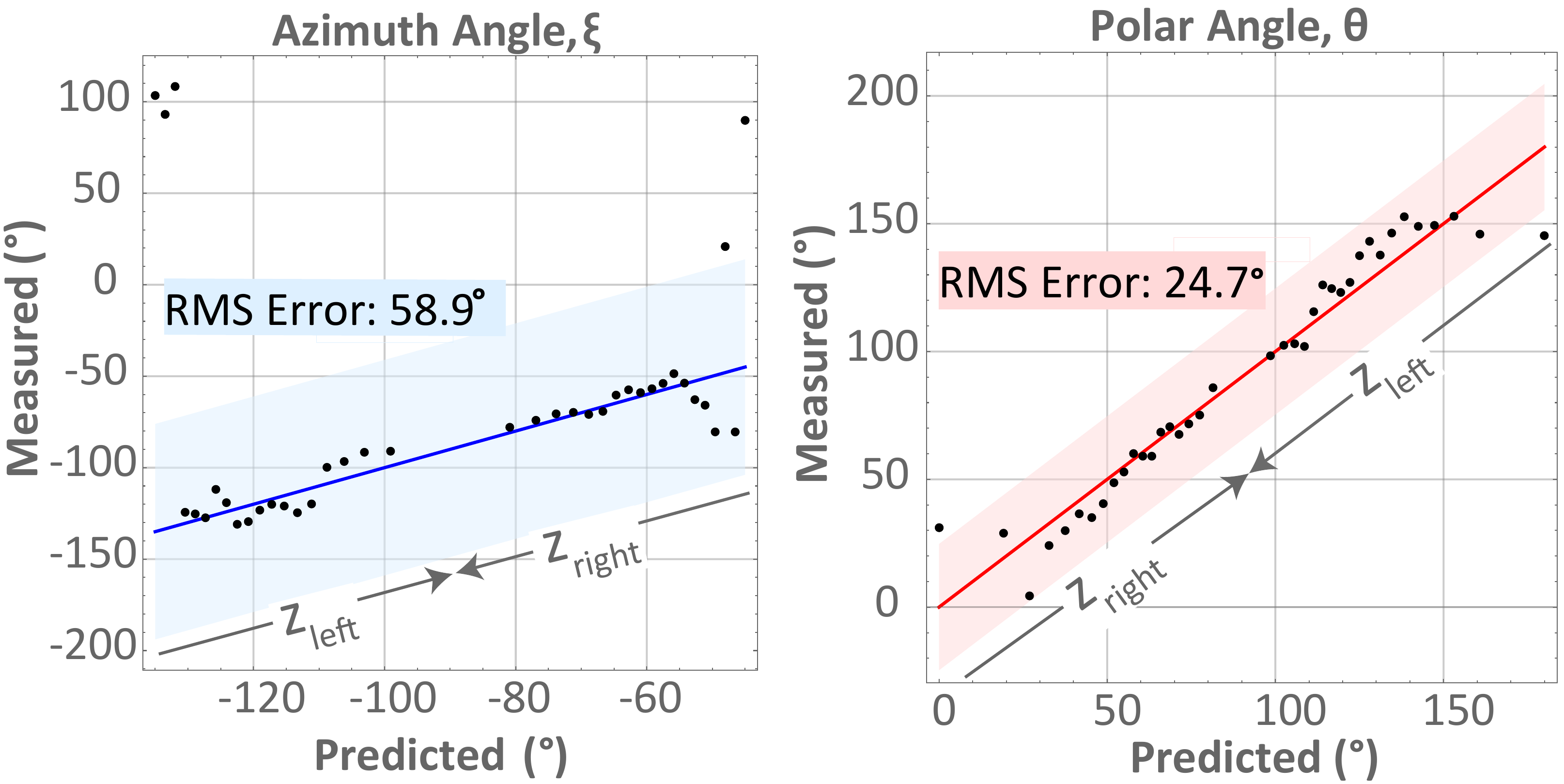}
	\end{center}
	\caption{\emph {Two vortex tilt evolution.} The left plot shows a comparison of the measured (dots) $\xi$ for the left and right vortices with propagation compared to the predicted (solid line) $\xi$ from Eq. (\ref{tilt_Indeb}) for each vortex. Similarly, the right plot shows the comparison of $\theta$ for both experiment and theory. The shaded regions show the rms error between the data and the predicted values. }
	\label{Indeb_Tilt_Comparison}
\end{figure}

%C:\Users\mlusk\Documents\Research\Topological_Fluids_of_Light\Mathematica\Indebetouw_Vortex_Pairs_Expt_Theory\Best_Linear_Core\Tilt_Tracker_Doublet_Annihilation_042520.nb
Turning attention to tilt, the evolution can be quantitatively measured from the experimental data by using the formalism developed in Sec. \ref{sec:tiltformalism} (see Appendix C). The experimentally measured tilts are directly compared with the analytical predictions in Fig. \ref{Indeb_Tilt_Comparison}. As the vortices evolve with propagation, it is clear that the vortex tilt not only is increasing until the annihilation point at which the orientations become equal, but closely follows the prediction, marked by the solid lines in the figure. The experimental measurements of both tilt angles are notably noisier near $z=0$, which we attribute not to lower-quality acquired data, but to lower sensitivity on both $\theta$ and $\xi$ as $\theta$ approaches 0 or $180^\circ$. For $\xi$, the orientation of the ellipse axes loses meaning for a circle, and for $\theta$, the change in ellipticity for a small change in $\theta$ is greatest near $\theta=90^\circ$ and approaches zero for $\theta$ approaching $0$ or $180^\circ$. This reduced sensitivity of tilt measurements for near-circular vortices was confirmed with synthetic data. With this in mind, we note that the fit between experiment and theory for tilt angles $\theta=90^\circ\pm 60^\circ$ is excellent.

%%%%%%%%%%%%%%%%%
\subsubsection{Confirming Tilt-Affected Vortex Hydrodynamics}
%%%%%%%%%%%%%%%%%

With the position and tilt evolution of a two-vortex system now well-established, we compare the results to the vortex kinematics of Eq. (\ref{v_final}) and alternative hydrodynamic approaches. For each approach, we calculate the velocity from the background fields at various $z$ steps according to Table \ref{table:1}. In the incompressible fluid case, the vortex velocity is governed purely by the background phase gradients, wherein the vortices are expected to propel each other forward along straight lines. In the compressible fluid case, vortex motion is determined by not only the phase gradients, but also the background amplitude gradients present.  For these incompressible and compressible cases, the background field is found by taking the expression for the paraxial field, Eq. (\ref{paraxial_Indeb}), and dividing out a circular vortex. In the final compressible plus tilt case, not only do background phase and amplitude gradients contribute to vortex motion, but the orientation of the vortex contributes to its motion, as it is a feature of the vortex itself. In order to find the background field for this case, the \emph{tilted} vortex,  described by Eqs. (\ref{tilt_Indeb}) and (\ref{position_Indeb}), is divided out from the expression for the paraxial field.

\begin{table}[b]
\centering
\begin{tabular}{|c | c | c |}
 \hline
 Fluid model  &  Velocity equation \\  [0.1ex]
 \hline
 \hline
 incompressible  &  ${\vec v} =\nabla \varphi_{bg}$ \\  [0.1ex]
 \hline
 compressible  &  ${\vec v} =  \nabla \varphi_{bg} - \hat{\kappa} \times \nabla \mathrm{ln} \rho_{bg} $  \\ [0.1ex] 
 \hline
 \hspace{0.1em} compressible plus tilt \hspace{0.1em} & \hspace{0.1em} ${\vec v} = \nabla_\perp \varphi_{bg} - \left(\VLCsq/\JLC\right)\pauli\nabla_\perp \mathrm{ln}\rho_{bg} $ \hspace{0.1em} \\ [0.1ex] 
 \hline
\end{tabular}
\caption{nondimensional models used to calculate velocities for Fig.~\ref{Vortex_Annihilation_Predicition_Comparisons_slide_3}~\cite{milne1996theoretical,Groszek2018}. In the compressible fluid velocity equation, $\hat{\kappa}$ is a unit vector in the direction of the vortex circulation vector for a circular vortex. For example, for a vortex of positive (negative) charge, $\hat{\kappa}=+\hat{z}$ ($\hat{\kappa}=-\hat{z}$).}
\label{table:1}
\end{table}

%
%FIGURE #11
%
\begin{figure}[t]
	\begin{center}
		\includegraphics[width=0.7\linewidth]{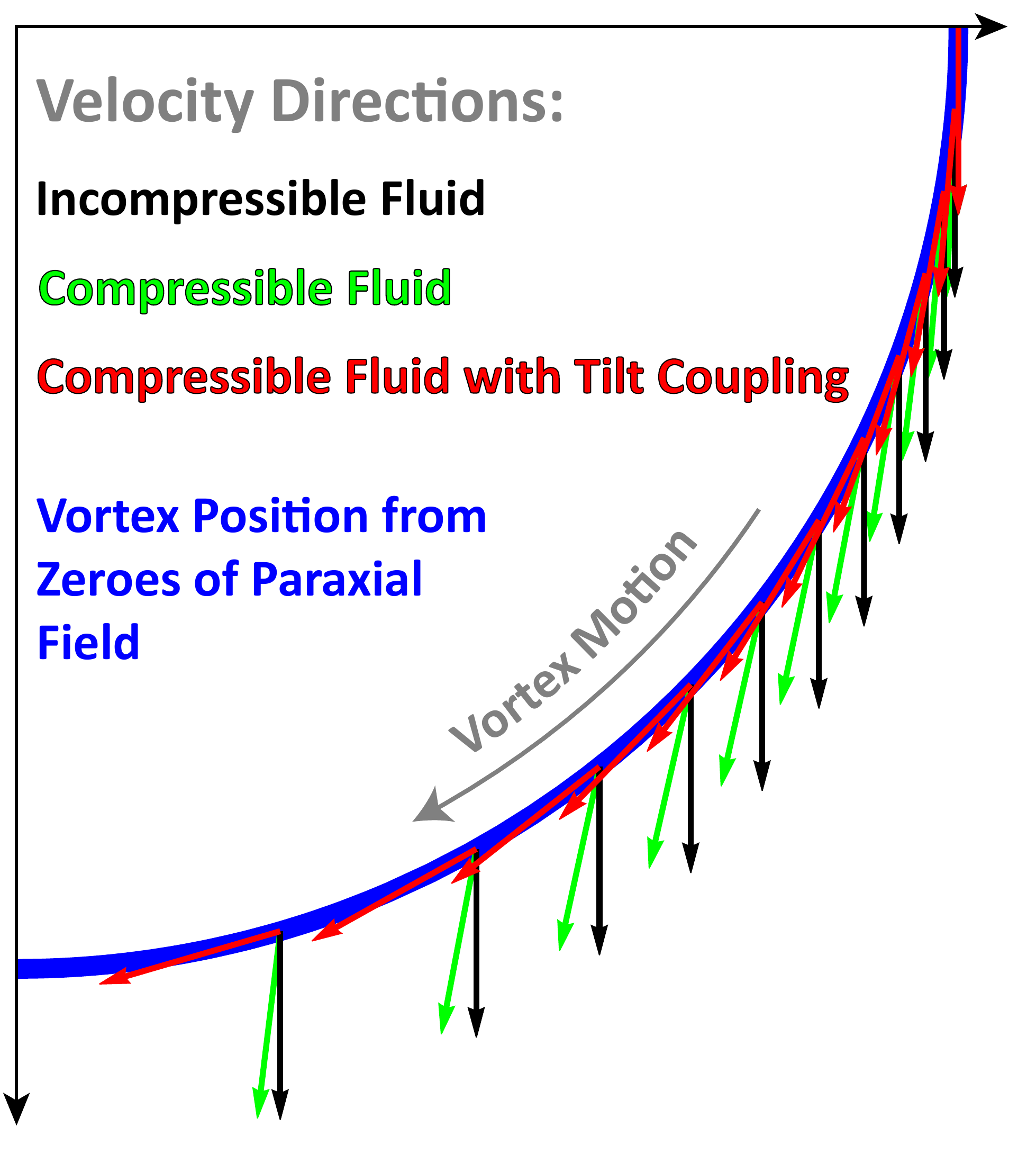}
	\end{center}
	\caption{\textit{Comparison of fluid models for predicting vortex pair motion}. The quarter-circle trajectory of a vortex in an opposite-pair annihilation event (blue solid line) with arrows shows the predicted velocities calculated from three different background fields, as described in the text and in Table I. The black incompressible fluid prediction (vertical arrows), addition of fluid compressibility in green (light gray arrows), and inclusion of vortex tilt in red (tangent arrows) are shown for the full kinetic theory of Eq. (21). Only the compressible fluid plus tilt prediction accurately describes the vortex motion. }
	\label{Vortex_Annihilation_Predicition_Comparisons_slide_3}
\end{figure}
%C:\Users\mlusk\Documents\Research\Topological_Fluids_of_Light\Mathematica\Indebetouw_Vortex_Pairs_Expt_Theory\Best_Linear_Core\Tilt_Tracker_Doublet_Annihilation_042520.nb

We compare each of these models to the known trajectory of one of the vortices in an opposite-charge pair, which follows a quarter-circle arc to recombination in the limit of an infinite beam waist (blue solid curve in Fig. \ref{Vortex_Annihilation_Predicition_Comparisons_slide_3}). The incompressible fluid prediction, [black (vertical) arrows in Fig. \ref{Vortex_Annihilation_Predicition_Comparisons_slide_3}], is clearly inadequate because it lacks the necessary transverse velocity component. Accounting for compressibility by including the background amplitude gradient, but not tilt [green (light gray) arrows in Fig. \ref{Vortex_Annihilation_Predicition_Comparisons_slide_3}], is a significant correction near $z=0$ where the vortices are mostly untilted, but it clearly breaks down near recombination where vortices are most tilted. Only if both the phase and amplitude gradients are included \textit{in addition to the evolving tilt of the vortex} [red tangent arrows in Fig. \ref{Vortex_Annihilation_Predicition_Comparisons_slide_3}, from Eq. (\ref{v_final})] do the velocity vectors fall tangent to the established trajectory. Clearly, vortex tilt plays an essential role in vortex motion in compressible hydrodynamics.\\

%%%%%%%%%%%%%%%%%%%%%%%%%%%%%%%%%%%%%%%%%%%%%%%%%%
\section{CONCLUSIONS}
%%%%%%%%%%%%%%%%%%%%%%%%%%%%%%%%%%%%%%%%%%%%%%%%%%
We have presented and experimentally verified a theory for predicting and understanding the dynamics of vortices in 2D fluids. Elliptic vortices are viewed as the projections of a construct, a circular vortex with an axis of symmetry that is tilted with respect to the propagation direction of the laser. This results in a more intuitive way of predicting and visualizing vortex-vortex dynamics. It is the coupling between this tilt and gradients in the background fluid density that was found to be essential in predicting vortex motion with a kinetic equation [Eq. (\ref{v_final})]. The coupling is not relevant for incompressible fluids that form the basis for much of our intuition, and it is not observed in compressible fluids for the many scenarios in which vortices are essentially rectilinear. However, it must be taken into account whenever vortex symmetry is lost, as is the case whenever vortices of opposite charge interact. In those settings, vortices exhibit the character of surfers since their orientation influences subsequent motion.

In the context of optical beams, it has long been known that both amplitude and phase gradients contribute to the motion of vortices~\cite{Rozas1999}, but a full hydrodynamic model for such systems had not yet been identified because the gradients alone are not sufficient. The inclusion of tilt in vortex kinematics is the key that unlocks a hydrodynamic interpretation of vortex motion in optical laser beams, even for purely linear systems  in which the vortex amplitude cores are always overlapping.

The coupling between tilt and background amplitude gradient has important implications for understanding and controlling the motion of vortices in other systems. While we derived the kinematics of tilted vortices in the context of propagating paraxial light, the results are equally valid for nonlinear 2D settings as well, including quantum fluids such as superfluid helium, atomic Bose-Einstein condensates, and nonlinear optical media.\\

\section{Acknowledgments}
The authors acknowledge interactions with L. Huzak and C. Zhu and support from the W.M. Keck Foundation and the NSF (Grant No. DMR 1553905).

%\bibliographystyle{prsty}
%\bibliography{tilt,AdditionalRefs,Supplemental}

\begin{thebibliography}{0}%
\makeatletter
\providecommand \@ifxundefined [1]{%
 \@ifx{#1\undefined}
}%
\providecommand \@ifnum [1]{%
 \ifnum #1\expandafter \@firstoftwo
 \else \expandafter \@secondoftwo
 \fi
}%
\providecommand \@ifx [1]{%
 \ifx #1\expandafter \@firstoftwo
 \else \expandafter \@secondoftwo
 \fi
}%
\providecommand \natexlab [1]{#1}%
\providecommand \enquote  [1]{``#1''}%
\providecommand \bibnamefont  [1]{#1}%
\providecommand \bibfnamefont [1]{#1}%
\providecommand \citenamefont [1]{#1}%
\providecommand \href@noop [0]{\@secondoftwo}%
\providecommand \href [0]{\begingroup \@sanitize@url \@href}%
\providecommand \@href[1]{\@@startlink{#1}\@@href}%
\providecommand \@@href[1]{\endgroup#1\@@endlink}%
\providecommand \@sanitize@url [0]{\catcode `\\12\catcode `\$12\catcode
  `\&12\catcode `\#12\catcode `\^12\catcode `\_12\catcode `\%12\relax}%
\providecommand \@@startlink[1]{}%
\providecommand \@@endlink[0]{}%
\providecommand \url  [0]{\begingroup\@sanitize@url \@url }%
\providecommand \@url [1]{\endgroup\@href {#1}{\urlprefix }}%
\providecommand \urlprefix  [0]{URL }%
\providecommand \Eprint [0]{\href }%
\providecommand \doibase [0]{https://doi.org/}%
\providecommand \selectlanguage [0]{\@gobble}%
\providecommand \bibinfo  [0]{\@secondoftwo}%
\providecommand \bibfield  [0]{\@secondoftwo}%
\providecommand \translation [1]{[#1]}%
\providecommand \BibitemOpen [0]{}%
\providecommand \bibitemStop [0]{}%
\providecommand \bibitemNoStop [0]{.\EOS\space}%
\providecommand \EOS [0]{\spacefactor3000\relax}%
\providecommand \BibitemShut  [1]{\csname bibitem#1\endcsname}%
\let\auto@bib@innerbib\@empty
%</preamble>
\end{thebibliography}%


\begin{thebibliography}{10}
	
	\bibitem{Mullins1964}
	W.~W. Mullins and R.~F. Sekerka, Journal of Applied Physics {\bf 35},  444
	(1964).
	
	\bibitem{Knowles1990}
	R. Abeyaratne and J.~K. Knowles, Journal of the Mechanics and Physics of Solids
	{\bf 38},  345  (1990).
	
	\bibitem{Hirth1998}
	J.~P. Hirth, H.~M. Zbib, and J. Lothe, Modelling and Simulation in Materials
	Science and Engineering {\bf 6},  165  (1998).
	
	\bibitem{Nilsen2006}
	H.~M. Nilsen, G. Baym, and C.~J. Pethick, Proceedings of the National Academy
	of Sciences {\bf 103},  7978  (2006).
	
	\bibitem{kivshar1998}
	Y.~S. Kivshar, J. Christou, V. Tikhonenko, B. Luther-Davies, and L.~M. Pismen, Optics Communications {\bf 152},
	198  (1998).
	
	\bibitem{Jezek2008}
	D.~M. Jezek and H.~M. Cataldo, Physical Review A {\bf 77},  043602  (2008).
	
	\bibitem{dosSantos2016}
	F.~E.~A. dos Santos, Physical Review A {\bf 94},  063633  (2016).
	
	\bibitem{Groszek2018}
	A.~J. Groszek, D.~M. Paganin, K. Helmerson, and T.~P. Simula, Physical Review A
	{\bf 97},  023617  (2018).
	
	\bibitem{Bershader1995}
	D. Bershader,  in {\em Fluid Vortices}, edited by S.~I. Green (Springer
	Netherlands, Dordrecht, 1995).
	
	\bibitem{Leweke2016}
	T. Leweke, S. Le~Dizès, and C.~H. Williamson, Annual Review of Fluid Mechanics
	{\bf 48},  507  (2016).
	
	\bibitem{kida1994}
	S. Kida and M. Takaoka, Annual Review of Fluid Mechanics {\bf 26},  169
	(1994).
	
	\bibitem{Eldredge2002}
	J.~D. Eldredge, T. Colonius, and A. Leonard, Journal of Computational Physics
	{\bf 179},  371   (2002).
	
	\bibitem{Allen1992}
	L. Allen, M.~W. Beijersbergen, R.~J.~C. Spreeuw, and J.~P. Woerdman, Physical
	Review A {\bf 45},  8185  (1992).
	
	\bibitem{Lax1975}
	M. Lax, W.~H. Louisell, and W.~B. McKnight, Physical Review A {\bf 11},  1365
	(1975).
	
	\bibitem{Madelung1927}
	E. Madelung, Zeitschrift f{\"u}r Physik {\bf 40},  322  (1927).
	
	\bibitem{Rozas1999}
	D. Rozas, Ph.D. thesis, Worcester Polytechnic Institute, 1999.
	
	\bibitem{Singh2003}
	R. Singh and S. Chowdhury, Optics Communications {\bf 215},  231  (2003).
	
	\bibitem{Bekshaev2003}
	A.~Y. Bekshaev, M.~S. Soskin, and M.~V. Vasnetsov, Journal of the Optical
	Society of America A {\bf 20},  1635  (2003).
	
	\bibitem{Zhao2017}
	P. Zhao, S. Li, Y. Wang, X. Feng, C. Kaiyu, L. Fang, W. Zhang, and Y. Huang,
	Scientific Reports {\bf 7},  7873  (2017).
	
	\bibitem{Kotlyar2017}
	V.~V. Kotlyar, A.~A. Kovalev, and A.~P. Porfirev, Physical Review A {\bf 95},
	053805  (2017).

	\bibitem{Indebetouw1993}
	G. Indebetouw, Journal of Modern Optics {\bf 40},  73  (1993).
	
	\bibitem{ChenRoux2008}
	M. Chen and F.~S. Roux, Journal of the Optical Society of America A {\bf 25},
	1279  (2008).
	
	\bibitem{andersen2019characterizing}
	J.~M. Andersen, S.~N. Alperin, A.~A. Voitiv, W.~G. Holtzmann, J.~T. Gopinath,
	and M.~E. Siemens, Applied optics {\bf 58},  404  (2019).
	
	\bibitem{Pethic}
	C. Pethick, {\em Bose-Einstein condensation in dilute gases} (Cambridge
	University Press, Cambridge New York, 2008).
	
	\bibitem{FourierFresnelPaper}
	T.~M. Pritchett and A.~D. Trubatch, American Journal of Physics {\bf 72},  1026
	(2004).
	
	\bibitem{gbur2016singular}
	G.~J. Gbur, {\em Singular optics} (CRC press, 2016).
	
	\bibitem{taylorerror}
	J.~R. Taylor, {\em Error Analysis} (Univ. Science Books, 1999).
	
	\bibitem{milne1996theoretical}
	 L.~M. Milne-Thomson, {\em Theoretical Hydrodynamics} (Courier Corporation, 1996).
	 
	\bibitem{Rozas1997}
	D. Rozas, Z.~S. Sacks, and G.~A. Swartzlander, Physical Review Letters {\bf
		79},  3399  (1997).
	
	\bibitem{gorshkov2002topology}
	V.~N. Gorshkov, A.~N. Kononenko, and M.~S. Soskin,  in {\em Selected Papers from Fifth International Conference on Correlation Optics}, edited by O. V. Angelesky, SPIE Proc. Vol 4607 (SPIE, Bellingham, 2002), pp. 13-24.
	
	\bibitem{huang2012low}
	D. Huang, H. Timmers, A. Roberts, N. Shivaram, and A.~S. Sandhu, American
	Journal of Physics {\bf 80},  211  (2012).

\end{thebibliography}

 \appendix
\setcounter{equation}{0}
% \counterwithin{figure}{section}
%%%%%%%%%%%%%
\section{Equipment Details}
%%%%%%%%%%%%%

The single-vortex and two-vortex experiments were each measured in independent setups. A diode laser, $\lambda=532$ nm, was used for the single-vortex experiments. For the two-vortex experiments, a $\lambda=633$ nm HeNe laser was used along with a longer stage. Both experiments use an Epson 83H projector LCD panel with panel dimensions of $9.52\times12.70$ mm$^2$ and pixel pitch of $12.4$ $\mu$ m as an SLM \cite{huang2012low}, and imaging lenses with $f=50$ cm. A Wincam LCM CCD was chosen because it is windowless, eliminating fringing from coverglass, and has a $2048\times2048$ resolution with pixel pitch of $5.5$ $\mu$ m.

%%%%%%%%%%%%%
\section{Phase-Shifting Digital Holography}
%%%%%%%%%%%%%

A hologram $\mathcal{H}$ to create a specific field is computer generated by summing a tilted plane wave and that field such that
\begin{eqnarray}
	\mathcal{H}&(x,y)& = \frac{1}{2} \bigg|e^{i \hspace{0.1em}\pi\left(\cos\alpha\hspace{0.02in} x + \sin\alpha\hspace{0.02in} y\right)/L} \nonumber \\
	&+&  A \psi_{\rm field}(x,y)+ B \psi_{\rm ref}(x,y,\phi_R)\bigg|,
	\label{eq:hologram}
\end{eqnarray}
where $L$ sets the grating spacing and $\alpha$ determines the angular orientation of the grating. In addition, $\psi_{field}$ is the field that is intended to be measured and $\psi_{ref}$ is an additional reference beam, only used when measuring the phase. Further, $A$ and $B$ are relative weights of the generated field and reference beams. For the single-vortex measurements, $\alpha=30^\circ$ was chosen so that the diffraction grating is not aligned with the pixels of the SLM to eliminate potential overlap of signal light with stray light from higher-order pixel diffraction, and $L=5$. In the two vortex experiments, using $\alpha=0$ and $L=10$ produced reasonable results.

To measure the full complex field, including both magnitude and phase, we use collinear phase-shifting digital holography \cite{andersen2019characterizing}. This technique uses five images generated from five unique holograms. One hologram with $A=1$ and $B=0$ (no reference is generated), such as one of those shown in Fig. 3(b)-(e), is used to obtain the intensity of the field, $I_{field}(x,y)$, from which the field magnitude is calculated as $\sqrt{I_{field}(x,y)}$. We divide the hologram by its maximum value such that all pixel values are between 0 and 1, to optimize the contrast on the SLM. The four remaining images, $I(x,y,\phi_R)$, used to reconstruct the phase, are the intensities of the field generated by four holograms in which the encoded beam is interfered with a dimensionless, collinear Gaussian reference with the same beam waist as $\psi_{field}(x,y)$ at unique phase shifts, $\phi_R = 0,\frac{\pi}{2},\pi,\frac{3\pi}{2}$. The relative weights $A$ and $B$ are chosen such that the reference power is low relative to the signal to better see optimal interference near a vortex core. Each of these four holograms is generated and then normalized to the maximum value of all four holograms. The measured intensities from the four interferograms can be used to calculate the phase at pixel location ($x,y$) via 
\begin{equation}
\Phi(x,y) = -\tan^{-1}\left(\frac{I(x,y,3\pi/2)-I(x,y,\pi/2)}{I(x,y,0)-I(x,y,\pi)}\right),
\end{equation} 
\noindent and $\psi_{meas}=\sqrt{I_{field}(x,y)} e^{i \Phi(x,y)}$ is the experimentally measured complex field.

The paraxial field of a single vortex, where $x_0$ and $w_0$ have units of distance, is
\begin{eqnarray}
	&&\psi_{\rm field, single}(x,y) =\frac{1}{w_0}\sqrt{\frac{2}{\pi}} [( (x - x_0) + i y \cos{\theta}]\cos{\xi} \nonumber\\
	&& +[y - i (x - x_0) \cos{\theta}] \sin{\xi})  e^{-(x^2+y^2)/w_0^2}.
\end{eqnarray}
This expression, with values chosen to work best within the constraints of the equipment, is used to generate the holograms for the single-vortex experiment with $A=0.9$ and $B= 0.1$. For the two-vortex experiment, an additional vortex is added with opposite sign such that the paraxial field is
\begin{eqnarray}
	\psi_{\rm field,pair}(x,y) &=& \frac{1}{w_0^2}\sqrt{\frac{2}{\pi}} ((x-x_0)+i y)\nonumber \\
	&& *((x+x_0)-i y)e^{-(x^2+y^2)/w_0^2}.
	\label{eq:twovortex}
\end{eqnarray}
In the two-vortex experiment, the nearly-overlapping vortex cores led to very low optical fields in the region of interest of the beam, so even lower reference power was used: $A=0.95$ and $B= 0.05$.

%%%%%%%%%%%%%%%%%%%%%%%%%%%%%%%%%%%
\section{Vortex Generation and Measurement}
%%%%%%%%%%%%%%%%%%%%%%%%%%%%%%%%%%%

%%%%%%%%
\subsection{Vortex Holograms} 
%%%%%%

%
%FIGURE # C1
%
\begin{figure}[b]
	\begin{center}
		\includegraphics[width=0.9\linewidth]{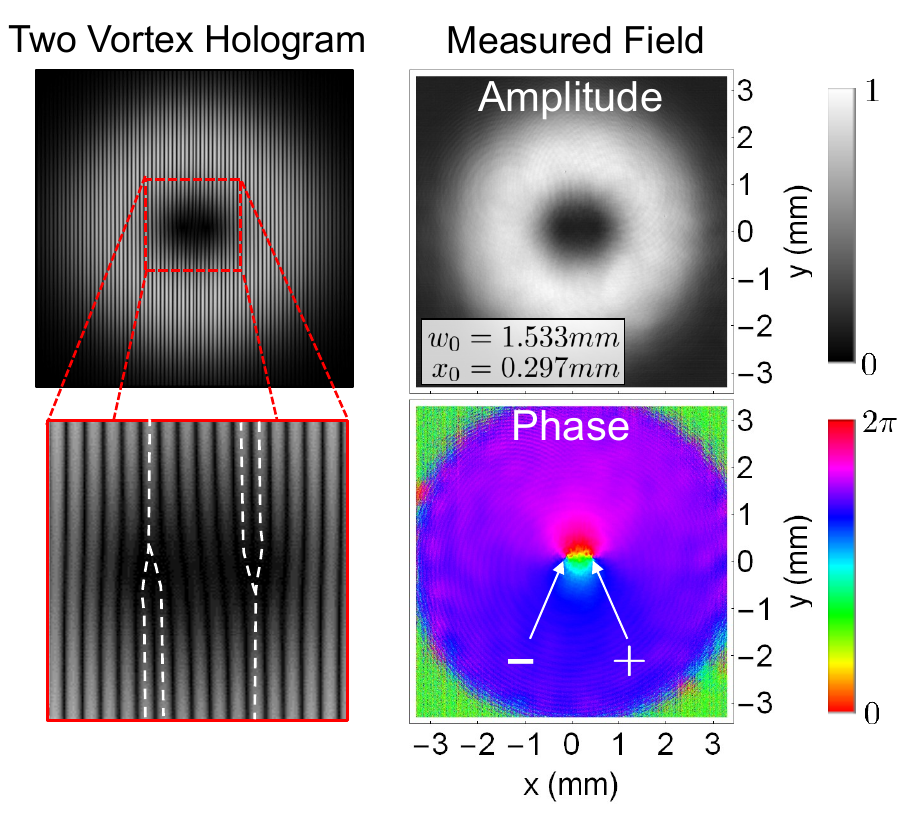}
	\end{center}
	\caption{ \emph{Vortex pair holograms and measured $z=0$ mm field.} The left column shows a vortex pair hologram with a close-up view of the vortex configuration. A white dashed outline traces over each fork to highlight them. The measured field amplitude and phase are shown in the right column. The inset on the amplitude plot shows the measured beam waist and vortex displacement, from a 2D amplitude fit using Eq. (\ref{eq:psi0twovortex}).}
	\label{fig:twovortexhologram}
\end{figure}
%%%% Experimental Data in Teams
% TFL -> Tilt (1- and 2-Vortex Repository -> Publication Data and Figures

The hologram, generated by substituting Eq. (\ref{eq:twovortex}) into Eq. (\ref{eq:hologram}), and the field measured at the $z=0$ imaging plane are shown in Fig. \ref{fig:twovortexhologram}. A close-up view of the vortex configuration is shown, along with white dashed outlines tracing over each fork to highlight the vortex locations in the hologram. The measured field amplitude and phase are shown in the right column. The inset on the amplitude plot shows the measured beam waist and vortex displacement, from a 2D amplitude fit using Eq. (\ref{eq:twovortex}).

%%%%%%%%
\subsection{Vortex Velocity} 
%%%%%%

\begin{figure}[hptb]
	\begin{center}
		\includegraphics[width=\linewidth]{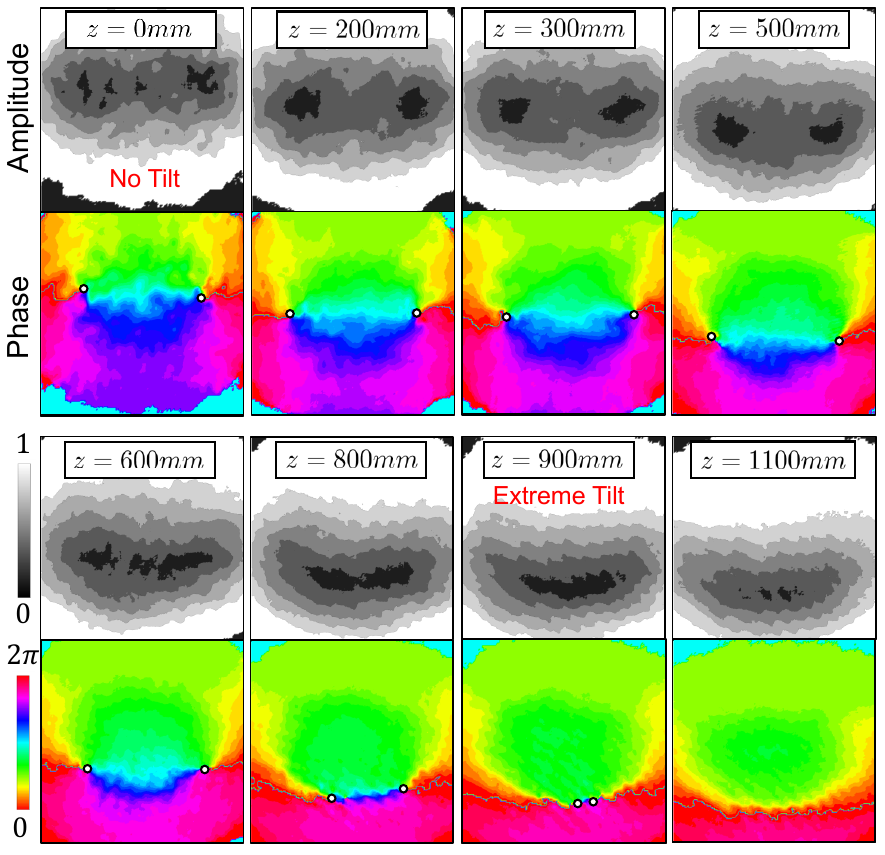}
	\end{center}
	\caption{ \emph{Snapshots of the measured amplitude and phase with propagation show the vortex trajectory and annihilation event.} Amplitude data are plotted with limited contours to highlight tilt evolution in the amplitude structure, but the data have resolution equal to that of the phase measurement. Data far beyond the annihilation point confirm that the vortices do not reemerge.
	}
	\label{fig:twovortexampphase}
\end{figure}

Measurements of the amplitude and phase evolution, plotted in Fig. \ref{fig:twovortexampphase}, show  close-ups of the vortex annihilation dynamics. The vortices are marked with white circles in the phase and move downward and towards each other until they annihilate each other and no vortices remain. Measurements following the annihilation event show no reappearance of the vortices. From the tracked vortex locations, the $y$ velocity of the vortices is measured. Using a linear fit on the combined data for both the left and right vortices, the measured $y$ velocity is $v_{y,meas}=-0.364\pm0.009$ mm/m as compared to the predicted $v_{y,model}=-0.314$ mm/m.

%
%FIGURE # C2
%

%%%%%%
\subsection{Vortex Tilt Measurements} 
%%%%%%
\label{Appendix:tiltmeas}
From the tilt formalism in Sec.II, we know that the gradient of $\psi$ is given by
$\nabla_\perp \psi = \nabla_\perp \textbf{F}^{-1} \textbf{r}_\perp = \textbf{F}^{-1}$. Using this relationship in conjunction with the relationship between $\Vsq$ and $\FLC$ from the polar decomposition theorem, tilt angles can be found via
\begin{equation}
	\textbf{V}^2  = \left[ \nabla_\perp \psi^{-1} \right] \left[ \nabla_\perp \psi^{-1} \right]^T.
	\label{vsqexp}
\end{equation}
\noindent where
\begin{eqnarray}
	\nabla_\perp \psi(x,y)&=&\nabla_\perp \begin{bmatrix} \mathrm{Re}[\psi(x,y)] \\
		\mathrm{Im}[\psi(x,y)]
	\end{bmatrix} \nonumber \\ 
&=& \begin{bmatrix} \partial_x \mathrm{Re}[\psi(x,y)] & \partial_y \mathrm{Re}[\psi(x,y)]  \\
		\partial_x \mathrm{Im}[\psi(x,y)] &  \partial_y \mathrm{Im}[\psi(x,y)]
	\end{bmatrix}.
	\label{eq:gradpsireim}
\end{eqnarray}
This shows that $\textbf{V}^2$ can be reconstructed in terms of the derivatives of the real and imaginary parts of the complex field at the location of the vortex for a given $z$ position, i.e., the full measured field can be used and the background field does not have to be calculated. 

The measured complex field is separated into its real and imaginary parts with vortex locations in the ($x,y$) plane found by the intersection of real and imaginary zeros. The $\textbf{V}^2$ matrix can be rewritten in terms of the gradient of an arbitrary field $\psi$. To obtain the slope at each vortex location, a small window is first cropped around the selected vortex. The real and imaginary parts are separately fit to planes of the form $z = a x +b y+c$.

\vskip 2 in

\end{document}

\title{Supplemental Information:\\Tilt Affects Vortex Dynamics in Beams of Light and Other Two-Dimensional, Compressible Fluids}  
	\author{Jasmine M. Andersen}
	\affiliation{Department of Physics and Astronomy, University of Denver, Denver, CO 80208, USA}
	\author{Andrew A. Voitiv}
	\affiliation{Department of Physics and Astronomy, University of Denver, Denver, CO 80208, USA}
	\author{Mark E. Siemens}
	\email{Mark.Siemens@du.edu}
	\affiliation{Department of Physics and Astronomy, University of Denver, Denver, CO 80208, USA}
	\author{Mark T. Lusk}
	\email{mlusk@mines.edu}
	\affiliation{Department of Physics, Colorado School of Mines, Golden, CO 80401, USA}

	\maketitle
%%%%%%%%%%%%%%%%%%%%%%%%%%%%%%%%%%%
\section{Theory Details}
%%%%%%%%%%%%%%%%%%%%%%%%%%%%%%%%%%%

%%%%%%%%%%%%%%%
\subsection{Visualization of Vortex Deformation by Matrix F}
\begin{figure}[hptb]
	\begin{center}
		\includegraphics[width=0.65\linewidth]{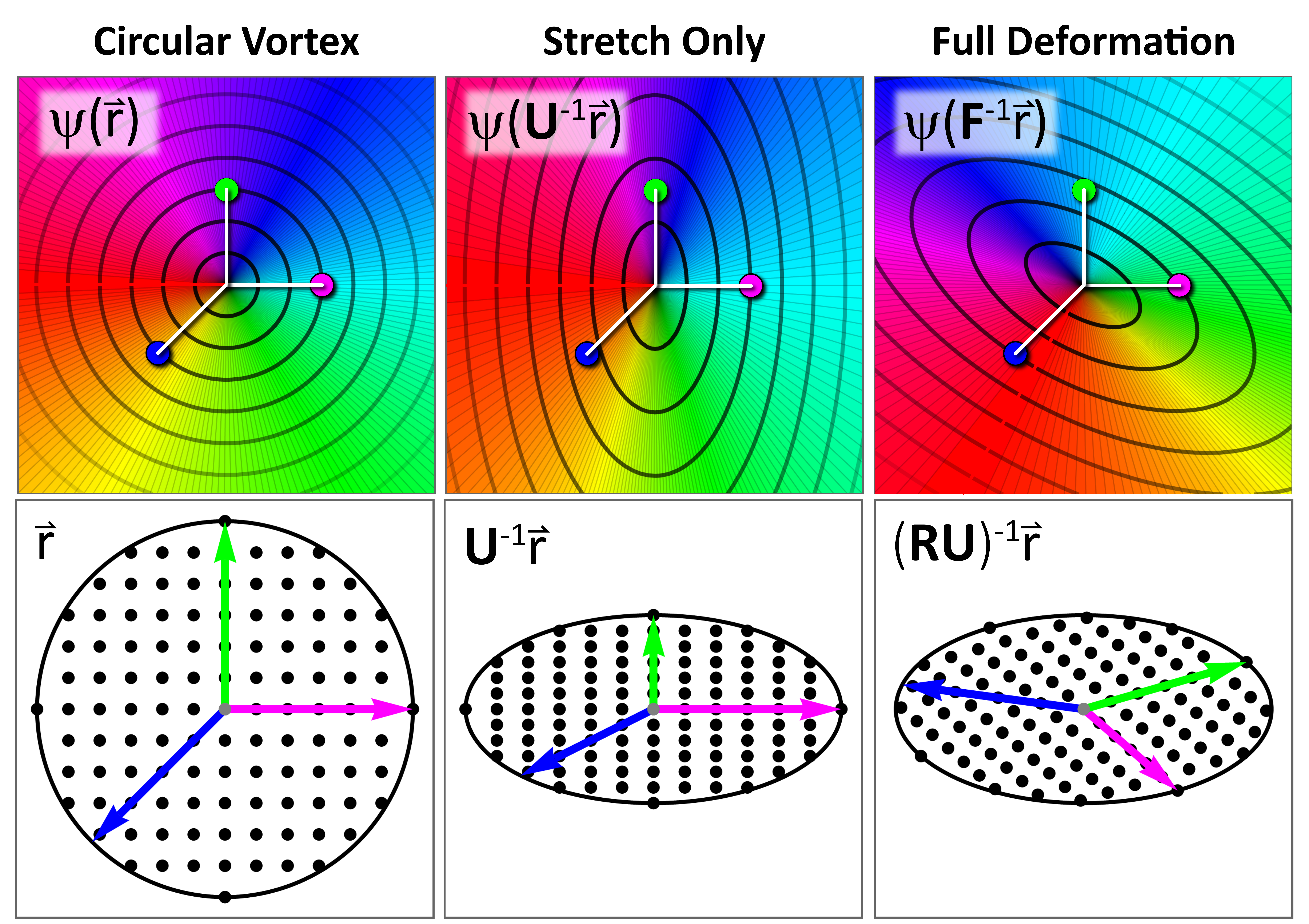}
	\end{center}
	\caption{\emph{ Homogeneous Deformation of a Circular Vortex}. (Top Rows) A 2D circular vortex, described in terms of its amplitude contours and phase, is vertically stretched and then rotated. Magenta, green, and blue markers identify the same positions in each panel. (Bottom Rows) The process can be viewed as the homogeneous deformation of a 2D continuum in which distance and direction, from the origin, are the magnitude and phase of $\psi$. Magenta, green, and blue arrows are the particular position vectors in the continuum that correspond to the colored markers in the top row.  Both tilt angles were chosen as $\xi=\theta=60^{\rm o}$ in this example.} 
	\label{Deformation_Primer}
\end{figure}

%%%%%%%%%%%%%%%
As shown in the top row of Fig. \ref{Deformation_Primer}, deformations can be viewed as vertically stretching an initially circular vortex by $\sec\theta$, followed by a counter-clockwise (CCW) rotation by an angle of $\xi$. An equivalent deformation perspective is shown in the bottom row, where the circular vortex is viewed as a reference configuration. 
%FIGURE #1
%
%
% Vortex_Tilt_2D_Continuum_Mechanics.nb

%%%%%%%%%%%%%%%
\subsection{Kinetic Equation and Laborious Reduction}
%%%%%%%%%%%%%%%

Consider a general paraxial vector field, $\vec u(x,y,z)$, that contains a single moving vortex with position described by ${\vecr}_v(z)$. Denote the vortex velocity by ${\vec v} = \partial_z{\vecr}_v$. Since $\vecu = \vec 0$ at the vortex center, its total derivative with respect to $z$ gives
\begin{equation}\label{v0}
d_z \vec u = \F^{-1}\vec v + \partial_z \vec u = \vec 0,
\end{equation}
with the transverse position evaluated at ${\vecr}_v(z)$ for all fields. A quick rearrangement of this result yields the vortex velocity as
\begin{equation}\label{v1}
{\vec v} = -\F \partial_z \vec u .
\end{equation} 
Re-write this using the paraxial equation
\begin{equation}\label{parax_vec}
	\pauli \partial_z \vec u = -\frac{1}{2} \nabla^2_\perp \vec u + {\cal V}\vec u,
\end{equation}
such that
\begin{equation}\label{parax_vec}
	{\vec v} = -\F \partial_z \vec u=  -\F \pauli^{-1} ( -\frac{1}{2} \nabla^2_\perp \vec u + {\cal V}\vec u).
\end{equation}
Then, we can rewrite F in terms of the background field instead, using $\textbf{F}$ and $\FLC$, $\F = \frac{1}{\rho_{bg}}\FLC\Phibg $ to obtain
\begin{equation}\label{v2}
{\vec v} = \frac{1}{2 \rho_{bg}} \FLC \Phibg\pauli^{-1} \nabla^2_\perp \vecu.
\end{equation}
where ${\cal V} \vec u=0$ since this is evaluated at the vortex center where $\vec u=0$.

This expression for vortex velocity can be simplified by noting that a vortex located at ${\vecr}_v$ is described by $\vecu_{\scriptscriptstyle v} = \FLCinv({\vecr}-{\vecr}_v)$ so $\vecu = \rho_{bg}\Phibg^T\FLCinv({\vecr}-{\vecr}_v)$. The Laplacian term of Eq. (\ref{v2}) evaluated at the vortex center, can therefore be written as
 \begin{equation}\label{lap}
 \nabla^2_\perp \vecu = 2 \nabla_\perp \cdot (\nabla_\perp \vecu )=2 \nabla_\perp \cdot (\rho_{bg} \Phibg^{-1} \FLC^{-1} ).
 \end{equation}
Substitution of this result into Eq. (\ref{v2}) gives
\begin{equation}\label{v2_2}
{\vec v} = \frac{1}{2 \rho_{bg}} \FLC \Phibg\pauli^{-1} \nabla^2_\perp 2 \nabla_\perp \cdot (\rho_{bg} \Phibg^{-1} \FLC^{-1} ).
\end{equation}

\noindent To simplify this, first consider a definition $\textbf{A} \equiv \FLC \Phibg$ so that $\vec v$ is then 
\begin{equation}
    \vec v= \frac{1}{\rho_{bg}}\textbf{A} \pauli \nabla_\perp \cdot (\rho_{bg} \textbf{A}^{-1})= \frac{1}{\rho_{bg}} \textbf{A} \pauli \left(  \rho_{bg} \nabla_\perp \cdot \textbf{A}^{-1} + \textbf{A}^{-1}\nabla_\perp \rho_{bg} \right).
\end{equation}
Since $\FLC$ is not a function of position in the $x$-$y$ plane, the first term can be reevaluated in terms of the gradient of the background phase only:
\begin{equation}
     \vec v= \frac{1}{\rho_{bg}} \textbf{A} \pauli \left(  \rho_{bg} (\nabla_\perp \Phibg^{-1}) \FLC^{-1} + \textbf{A}^{-1}\nabla_\perp \rho_{bg} \right).
\end{equation}

\noindent Now, we can expand to two terms and simplify using $\left(\nabla_\perp \rho_{bg}\right)/\rho_{bg}= \nabla_\perp log(\rho_{bg})$ and the velocity becomes 
\begin{equation}
     \vec v= \textbf{A} \pauli (\nabla_\perp \Phibg^{-1}) \FLC^{-1} + \textbf{A} \pauli \textbf{A}^{-1}\nabla_\perp \log \rho_{bg} .
\end{equation}

\noindent For clarity, we will rewrite this expression as 
\begin{equation}
    \vec v = \vec v_{\varphi}+ \vec v_{\xi,\theta,\rho}
\end{equation}

Next, we simplify the phase contribution, $\vec v_{\varphi}=\textbf{A} \pauli (\nabla_\perp \Phibg^{-1}) \FLC^{-1}$, to the vortex velocity. The three tensor,
\begin{equation}
   \nabla_\perp \Phibg^{-1}=
        \begin{bmatrix}
            \begin{pmatrix}
            -\sin \varphi_{bg} \partial_x \varphi_{bg} \\
            -\sin \varphi_{bg} \partial_y \varphi_{bg} 
            \end{pmatrix} & 
            \begin{pmatrix}
            -\cos\varphi_{bg} \partial_x \varphi_{bg} \\
           -\cos\varphi_{bg} \partial_y \varphi_{bg} 
            \end{pmatrix} \\
            \begin{pmatrix}
           \cos\varphi_{bg} \partial_x \varphi_{bg} \\
            \cos \varphi_{bg} \partial_y \varphi_{bg} 
            \end{pmatrix} & 
            \begin{pmatrix}
            -\sin\varphi_{bg} \partial_x \varphi_{bg} \\
            -\sin\varphi_{bg} \partial_y \varphi_{bg} 
            \end{pmatrix} 
        \end{bmatrix}
\end{equation}
\noindent acts on $\FLC$ to produce a vector

\begin{equation}
     \left(\nabla_\perp \Phibg^{-1}\right) \FLC^{-1}=        
     \begin{bmatrix}
            -\left(\cos\theta \hspace{0.01in} \cos\xi \hspace{0.01in} \cos\varphi_{bg}+\sin\xi \hspace{0.01in}\sin\varphi_{bg} \right) \partial_y \varphi_{bg}+\left(\cos\theta\hspace{0.01in} \cos\varphi_{bg}\hspace{0.01in} \sin\xi-\cos\xi \hspace{0.01in}\sin\varphi_{bg} \right) \partial_x \varphi_{bg}
            \\
            \left(\cos\varphi_{bg}\hspace{0.01in} \sin\xi-\cos\theta\hspace{0.01in} \cos\xi \hspace{0.01in}\sin\varphi_{bg} \right) \partial_y\varphi_{bg}+\left(\cos\xi \hspace{0.01in}\cos\varphi_{bg}+\cos\theta \hspace{0.01in}\sin\xi\hspace{0.01in} \sin\varphi_{bg} \right)\partial_x\varphi_{bg}
        \end{bmatrix}.
\end{equation}
 \noindent This vector can now be used in simplifying the entire first term for which the result is
 \begin{equation}
      \vec v_{\varphi}=-\textbf{A} \pauli \left(\nabla_\perp \Phibg^{-1}\right) \FLC^{-1}= 
      \begin{bmatrix}
            \partial_x\varphi_{bg}
            \\
            \partial_y\varphi_{bg}
      \end{bmatrix}
      =\nabla_\perp \varphi_{bg}
      \label{eq:SIvrho}
 \end{equation}
 
 \noindent as expected. 
 
Next, the second term, $\vec v_{\xi,\theta,\rho}=\textbf{A} \pauli \textbf{A}^{-1}\nabla_\perp \log \rho_{bg}$, is directly compared to the result from Eq. (20) of the manuscript, $\vec v_{\xi,\theta,\rho}=- \frac{\VLCsq}{\JLC}\pauli\nabla_\perp \log\rho_{bg}$ to verify their equivalence. The derived expression of \ref{eq:SIvrho} when evaluated is given by 

\begin{equation}
    v_{\xi,\theta,\rho}=\textbf{A} \pauli \textbf{A}^{-1}\nabla_\perp \log \rho_{bg}=
        \begin{bmatrix}
            -\cos\xi \hspace{0.01in} \sin\theta \hspace{0.01in} \sin\xi \hspace{0.01in} \tan\theta & -\sec\theta \left( \cos^2\theta \hspace{0.01in} \cos^2\xi+\sin^2\xi \right) \\
            \cos^2\xi\hspace{0.01in} \sec\theta + \cos\theta\hspace{0.01in} \sin^2\xi & \cos\xi \hspace{0.01in}\sin\theta\hspace{0.01in} \sin\xi\hspace{0.01in} \tan\theta
        \end{bmatrix}   
        \nabla_\perp \log \rho_{bg}.
\end{equation}

\noindent Similarly, the result in the manuscript when evaluated is given by 

\begin{equation}
    v_{\xi,\theta,\rho}=- \frac{\VLCsq}{\JLC}\pauli\nabla_\perp \log\rho_{bg}=
        \begin{bmatrix}
            -\cos\xi \hspace{0.01in} \sin\theta \hspace{0.01in} \sin\xi \hspace{0.01in} \tan\theta & -\cos\theta \left( \cos^2\xi +\sec^2\theta \hspace{0.01in}\sin^2\xi \right) \\
            \cos^2\xi\hspace{0.01in} \sec\theta + \cos\theta\hspace{0.01in} \sin^2\xi & \cos\xi \hspace{0.01in}\sin\theta\hspace{0.01in} \sin\xi\hspace{0.01in} \tan\theta
        \end{bmatrix}   
        \nabla_\perp \log \rho_{bg}.
\end{equation}
These two terms, when subtracted yield the zero matrix and are therefore equivalent. The final velocity, as in Eq. (20) of the manuscript, is then written as
\begin{equation}\label{v_final}
{\vec v} = \nabla_\perp \varphi_{bg} - \frac{\VLCsq}{\JLC}\pauli\nabla_\perp \log\rho_{bg} .
\end{equation}
%Rings_Fields_2D_Projections_042520.nb
where $\JLC = {\rm Det}(\FLC)$.

%%%%%%%%%%%%%%%
\subsection{Fourier Transform Method for Analytical Fresnel Integration}

The propagation of optical fields with vortices can be calculated with the Fresnel integral. We analytically implement this calculation with Fourier transforms, following  Pritchett and Trubatch \cite{FourierFresnelPaper}.

\subsubsection{Basic setup for the calculation}
We want to solve the following system:
\begin{equation}
    \partial_{xx} u + \partial_{yy}  u + 2 k \imath \partial_z  u = 0,
\end{equation}
which is the paraxial equation for complex amplitude
\begin{equation}
     u(x,y,z=0) =  u_{Gauss}(x,y,0) \tilde u (x,y),
\end{equation}
where $u_{Gauss}(x,y,0)$ represents the complex-valued initial background mode, a Gaussian function for these purposes, which is modulated by the amplitude and phase of one or more vortices described by $\tilde u(x,y)$.

The solution to this system is the following Fresnel integral, a convolution of the initial condition with the paraxial Green function:
\begin{equation}
    u(x,y,z) = \frac{-\imath k}{z} \int dx' \int dy' u(x', y', 0) e^{\frac{\imath k}{2z} \left( (x-x')^2+(y-y')^2\right)}.
\end{equation}
Non-dimensionalize the paraxial equation using:
\begin{equation}
    x \rightarrow w_0 x, y \rightarrow w_0 y, z \rightarrow k w_0^2 z. 
\end{equation}
The non-dimensional paraxial equation is then
\begin{equation}
    \partial_{xx} u + \partial_{yy} u + 2 \imath \partial_z u = 0
\end{equation}
and the non-dimensional Fresnel integral is:
\begin{equation}
    u(x,y,z) = \frac{-\imath}{z} \int dx_1 \int dy_1 u(x_1, y_1, 0) e^{\frac{\imath}{2z} \left( (x-x_1)^2+(y-y_1)^2\right)}.
\end{equation}

Then the Fresnel integral is
\begin{equation}
    u(x,y,z) = \frac{-\imath}{z}\frac{1}{2\pi} \int dx_1 \int dy_1 \frac{2}{\sqrt{\pi}} e^{-(x_1^2+y_1^2)} \tilde u (x_1,y_1)e^{\frac{\imath}{2z}\left((x-x_1)^2 + (y-y_1)^2 \right)}.
\end{equation}
We can write this with the $x_1$ and $y_1$ terms grouped by integral:
\begin{equation}
    u(x,y,z) = \frac{-\imath}{z}\frac{1}{2\pi}\frac{2}{\sqrt \pi} \int dx_1 e^{-x_1^2 (1-\frac{\imath}{2 z})} e^{-\imath x_1 \frac{x}{z}} \int dy_1 \tilde u (x_1,y_1) e^{-y_1^2 (1-\frac{\imath}{2 z})} e^{-\imath y_1 \frac{y}{z}}.
\end{equation}

Now define the following:
\begin{equation}
    \alpha = 1-\frac{\imath}{2 z}, \beta_x = \frac{-x}{z}, \beta_y = \frac{-y}{z}.
\end{equation}
Then the non-dimensional Fresnel integral is
\begin{equation}
    u(x,y,z) = \frac{-\imath}{\pi^{3/2}} e^{\frac{\imath}{2z} (x^2+y^2)} \int dx_1 e^{-\alpha x_1^2} e^{\imath \beta_x x_1} \int dy_1 \tilde u (x_1,y_1) e^{-\alpha y_1^2} e^{\imath \beta_y y_1},
\end{equation}
which is in the form of a two-dimensional Fourier transform. Define the Fourier transform as follows:
\begin{equation}\label{Fourier}
    F_q(\beta_q, q) = \frac{1}{\sqrt{2 \pi}} \int dq f_q(q) e^{\imath \beta_q q},
\end{equation}
where $q$ represents $x$ or $y$, and define the following kernel functions:
\begin{equation} \label{FTx}
f_x(x_1) = \tilde u(x_1,y_1) e^{-\alpha x_1^2}
\end{equation}
\begin{equation}\label{FTy}
f_y(y_1) = e^{-\alpha y_1^2} F_x(\beta_x;x_1).
\end{equation}

With this, the Fresnel integral becomes
\begin{equation} \label{Result}
    u(x,y,z) = \frac{-\imath}{z}{2}{\sqrt \pi} e^{\imath \frac{x^2+y^2}{2 z}}F_y(\beta_x; \beta_y)
\end{equation}
Parameters listed after semicolons serve as reminders of functional dependence.

\subsubsection{Implementation for single titled vortex on the shoulder of a Gaussian beam}
We take the initial field
\begin{equation} \label{Single}
    \tilde u(x,y) = ((x - x_0) + \imath (y - y_0) \cos{\theta})\cos{\xi} + ((y - y_0) - \imath (x - x_0) \cos{\theta}) \sin{\xi},
\end{equation}
to be a single vortex that may be tilted at any initial angles $\xi$ and $\theta$ and which may be shifted along the $x$- or $y$-axes by $x_0$ and $y_0$, respectively. For this demonstration, we set: $y_0 = 0$.

To reach the desired field, we begin by inserting Equation \ref{Single} into Equation \ref{FTx} and then using Equation \ref{Fourier}:
\begin{equation}
    F_x(\beta_x) = \frac{1}{\sqrt{2 \pi}} \int dx f_x(x_1) e^{\imath \beta_x x}.
\end{equation}
The result of this Fourier transform is used to construct the next (using Equation \ref{FTy}):
\begin{equation}
    F_y(\beta_y) = \frac{1}{\sqrt{2 \pi}} \int dy f_y(y_1) e^{\imath \beta_y y}.
\end{equation}

The final Fresnel-integrated field is thus evaluated by finding (Equation \ref{Result}):
\begin{equation}
    u(x,y,z) = \frac{-\imath}{z}{2}{\sqrt \pi} e^{\imath \frac{x^2+y^2}{2 z}} F_y(\beta_y),
\end{equation}
with result:
\begin{equation}
    u(x,y,z)=
    \sqrt{\frac{2}{\pi }} e^{\frac{\imath \left(x^2+y^2\right)}{2 (z-\imath)}}
    \frac{ (\cos
   \text{$\xi $} (-x+\imath x_0 z+x_0-\imath y \cos \text{$\theta $})+\sin
   \text{$\xi $} (-y+\cos \text{$\theta $} (x_0 z+\imath
   (x-x_0))))}{(z-\imath)^2}
   \label{eq:SVGDimensionalFresnel}
\end{equation}

\subsubsection{Implementation for a vortex pair on a Gaussian}
We take our vortex pair to consist of opposite unit charge vortices with linear amplitude cores, $(x + \imath y)$ and $(x - \imath y)$, which are neither initially tilted but are each shifted from the origin by $\pm x_0$:
\begin{equation}
    \tilde u(x,y) = ((x + x_0) - \imath y) \times ((x - x_0) + \imath y).
\end{equation}

The order of calculations follows directly the order used in the previous implementation for the single-vortex case, for the new expression of $\tilde u(x,y)$. The final Fresnel-integrated result is:
\begin{equation}
    u(x,y,z) = 
    \sqrt{\frac{2}{\pi }} e^{\frac{\imath \left(x^2+y^2\right)}{2 (z-\imath)}}
    \frac{\left(x^2+x_0^2 (z-\imath)^2-2 x_0 y (z-\imath)+y^2-2 z (z-\imath)\right)}{(1+\imath z)^3}
\end{equation}

%%%%%%%%%%%%%%%
\subsection{Two Vortex Evolution}
\begin{figure}[htbp]
	\begin{center}
		\includegraphics[width=0.4\linewidth]{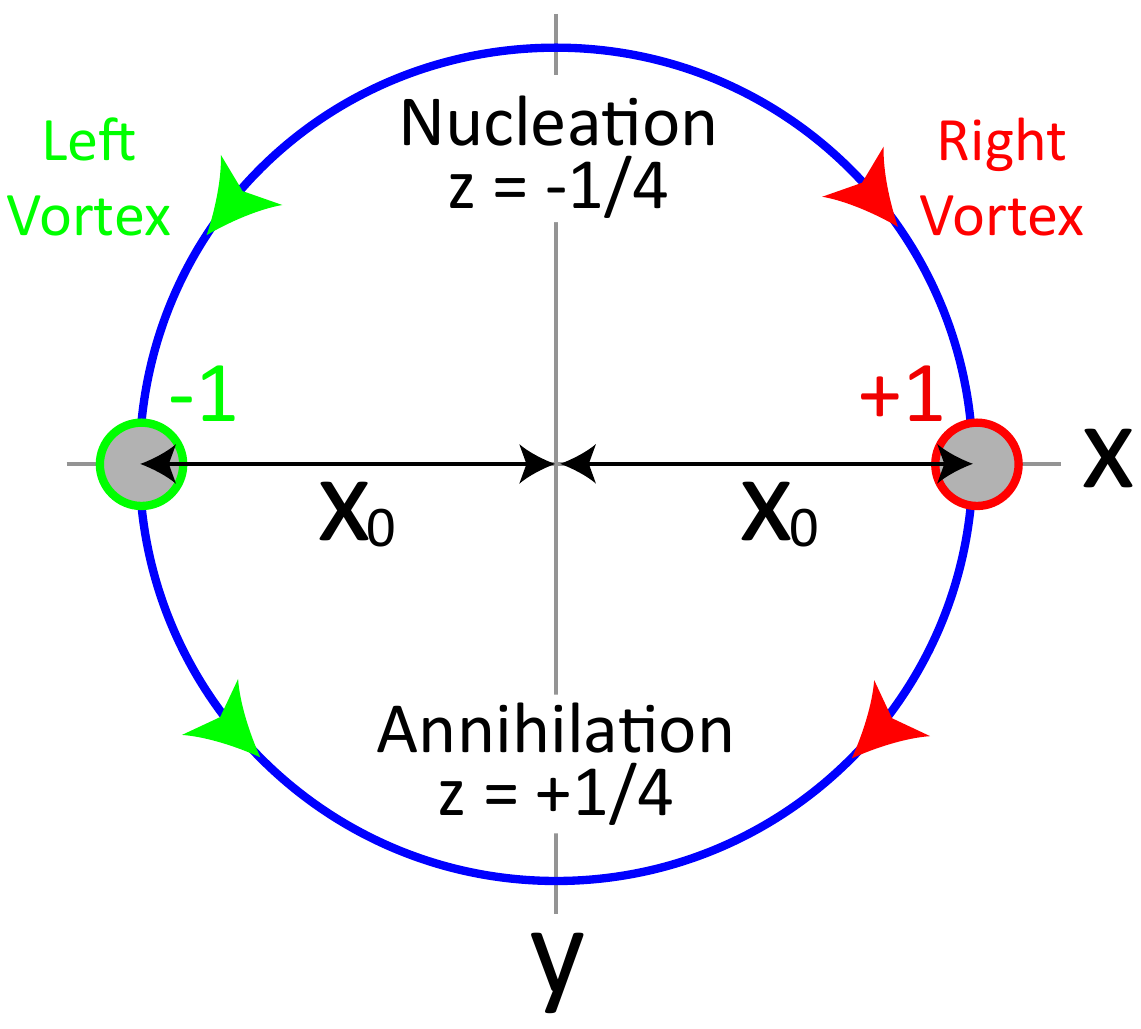}
	\end{center}
	\caption{ \textit{Trajectory of an Oppositely-Charged Vortex Pair, as Described by Eq. (26) of the Manuscript}.}
	\label{Indeb_0}
\end{figure}

In the infinite beam case, as shown in Fig. \ref{Indeb_0}, a vortex pair that nucleates at $z = -1/4$ annihilates at $z = +1/4$, and traces out a circular trajectory in the x,y-plane, which are consistent results with work done by Indebetouw~\cite{Indebetouw}. For finite beams, a very nearly circular trajectory is expected for $x_0 \ll w_0$.
%
%FIGURE 8 #
%
%
% Vortex_Tilt_Helicity\Momentum_Flux_Indebetouw_012520.nb

\begin{figure}[hptb]
	\begin{center}
		\includegraphics[width=0.6\textwidth]{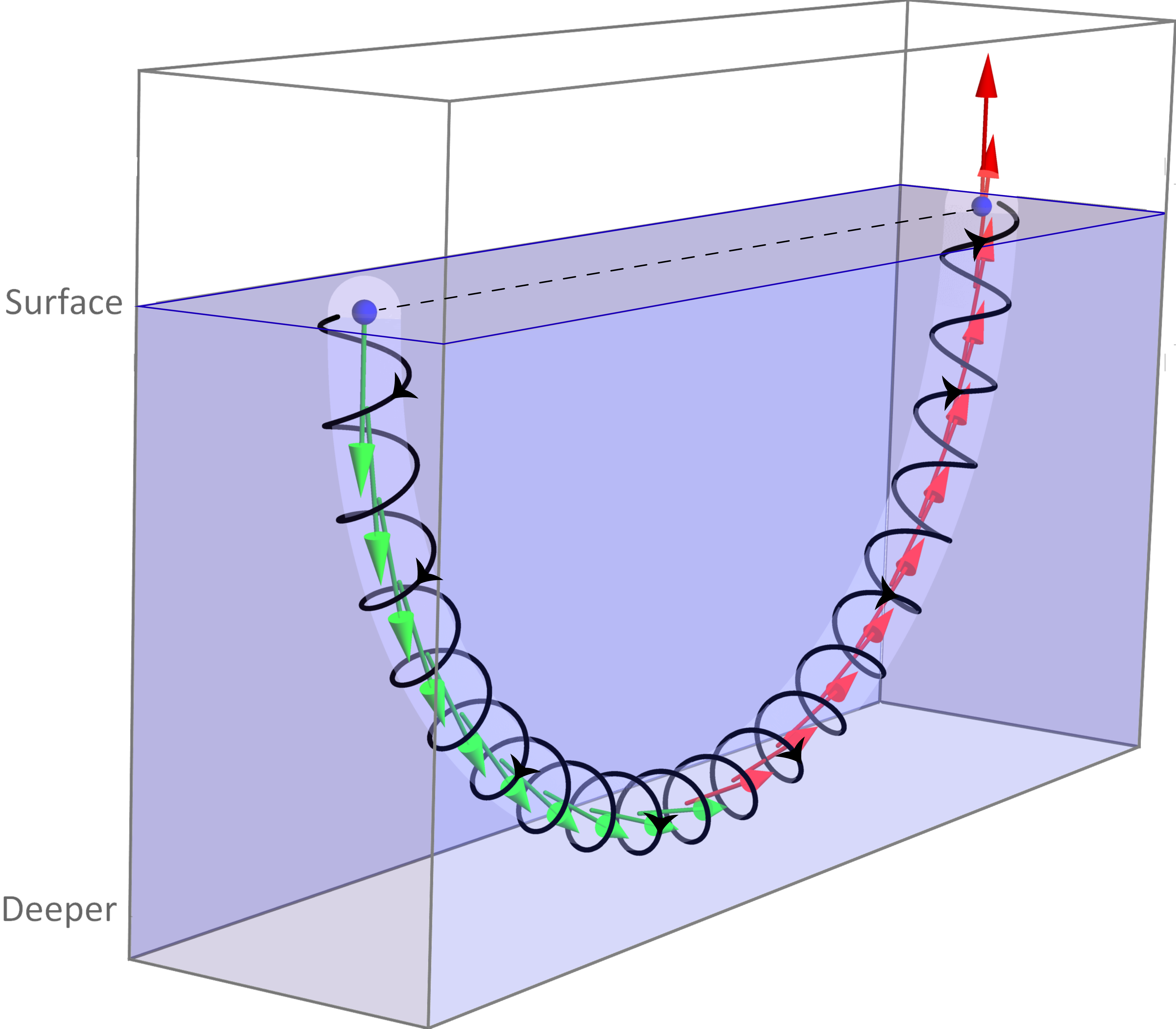}
	\end{center}
	\caption{ \emph {Hydrodynamic Vortex Half-loop.} Vortex half-loops can be produced in water, and water spins in opposite directions at each end. This is analogous to the optical dipole in which the vortex charges are of opposite sign at z = 0.}
	\label{water_vortex}
\end{figure}

The vortex loop of Fig. 10 in the manuscript is reminiscent of the half-loop vortices observed in water, shown in Fig. \ref{water_vortex}. In both cases, there is a continuous transition of vortex charge.
%
%FIGURE 11 #
%
%
% Momentum_Flux_Indebetouw_052020.nb

%%%%%%%%%%%%%%%

%%%%%%%%%%%%%%%%%%%%%%%%%%%%%%%%%%%
\section{Experiment and Measurement Details}
%%%%%%%%%%%%%%%%%%%%%%%%%%%%%%%%%%%
%%%%%%%%%%%%%
\subsection{Experimental Apparatus Details}
%%%%%%%%%%%%%

The single vortex and two vortex experiments were each measured in independent setups. A $\lambda=532$ nm diode laser was used for the single vortex experiments. For the two vortex experiments, a $\lambda=633$ nm HeNe laser was used along with a longer stage. Both experiments use an Epson 83H projector LCD panel with panel dimensions of $9.52$ mm by $12.70$ mm and pixel pitch of $12.4 \mu m$ as an SLM \cite{huang2012low}, and imaging lenses with $f=50$ cm. A Wincam LCM CCD was chosen because it is windowless, eliminating fringing from coverglass, and has a 2048x2048 resolution with pixel pitch of $5.5 \mu m$.

%%%%%%%%%%%%%
\subsection{Phase-shifting Digital Holography  Details}
%%%%%%%%%%%%%

A hologram to create a specific field is computer generated by summing a tilted plane wave and that field such that
\begin{equation}
    \mathrm{Hologram}(x,y) = \frac{1}{2} \left|  e^{\imath \hspace{0.1em}\pi\left(\cos\alpha\hspace{0.02in} x + \sin\alpha\hspace{0.02in} y\right)/L} +  A \psi_{field}(x,y)+ B \psi_{ref}(x,y,\phi_R)\right|,
    \label{eq:hologram}
\end{equation}
where $L$ sets the grating spacing and $\alpha$ determines the angular orientation of the grating. $\psi_{field}(x,y)$ is the field that is intended to be measured and $\psi_{ref}(x,y,\phi_R)$ is an additional reference beam, only used when measuring the phase. $A$ and $B$ are relative weights of the generated field and reference beams. For the single vortex measurements, $\alpha=30^\circ$ was chosen so that the diffraction grating is not aligned with the pixels of the SLM to eliminate potential overlap of signal light with stray light from higher order pixel diffraction, and $L=5$. In the two vortex experiments, using $\alpha=0$ and $L=10$ produced reasonable results.

To measure the full complex field, including both magnitude and phase, we use collinear phase-shifting digital holography \cite{andersen2019characterizing}. This technique uses five images generated from five unique holograms. One hologram with $A=1$ and $B=0$ (no reference is generated), such as one of those shown in Fig. 3 (b-e) of the manuscript, is used to obtain the intensity of the field, $I_{field}(x,y)$, from which the field magnitude is calculated as $\sqrt{I_{field}(x,y)}$. We divide the hologram by its maximum value such that all pixel values are between 0 and 1, to optimize the contrast on the SLM. The four remaining images, $I(x,y,\phi_R)$, used to reconstruct the phase, are the intensities of the field generated by four holograms in which the encoded beam is interfered with a dimensionless, collinear Gaussian reference with the same beam waist as $\psi_{field}(x,y)$ at unique phase shifts, $\phi_R = 0,\frac{\pi}{2},\pi,\frac{3\pi}{2}$. $A$ and $B$ are chosen such that the reference power is low relative to the signal to better see optimal interference near a vortex core. Each of these four holograms are generated and then normalized to the maximum value of all four holograms. The measured intensities from the four interferograms can be used to calculate the phase at pixel location ($x,y$) via 
\begin{equation}
    \Phi(x,y) = -\tan^{-1}\left(\frac{I(x,y,\phi_R=3\pi/2)-I(x,y,\phi_R=\pi/2)}{I(x,y,\phi_R=0)-I(x,y,\phi_R=\pi))}\right),
\end{equation} 
\noindent and $\psi_{meas}=\sqrt{I_{field}(x,y)} e^{\imath \Phi(x,y)}$ is the experimentally measured complex field.

The paraxial field of a single vortex, where $x_0$ and $w_0$ have units of distance, is
\begin{equation}
    \psi_{\rm field, single}(x,y) =\frac{1}{w_0}\sqrt{\frac{2}{\pi}}\left( (x - x_0) + \imath y \cos{\theta})\cos{\xi}\\
    +(y - \imath (x - x_0) \cos{\theta}) \sin{\xi} \right) e^{-(x^2+y^2)/w_0^2}.
        \label{}
\end{equation}
This expression, with values chosen to work best within the constraints of the equipment, is used to generate the holograms for the single vortex experiment with $A=0.9$ and $B= 0.1$. For the two-vortex experiment, an additional vortex is added with opposite sign such that the paraxial field is
\begin{equation}
\psi_{\rm field,pair}(x,y) = \frac{1}{w_0^2}\sqrt{\frac{2}{\pi}} ((x-x_0)+\imath y) ((x+x_0)-\imath y) e^{-(x^2+y^2)/w_0^2}.
        \label{eq:twovortex}
\end{equation}
In the two-vortex experiment, the nearly-overlapping vortex cores led to very low optical fields in the region of interest of the beam, so even smaller reference power was used: $A=0.95$ and $B= 0.05$.

%%%%%%%%%%%%%%%%%%%%%%%%%%%%%%%%%%%
\subsection{Single Vortex Measurements}
%%%%%%%%%%%%%%%%%%%%%%%%%%%%%%%%%%%

%%%%%%%%%%%%%
\subsubsection{Vortex Velocity Measurements}
\begin{figure}[h]
\includegraphics[width=\textwidth]{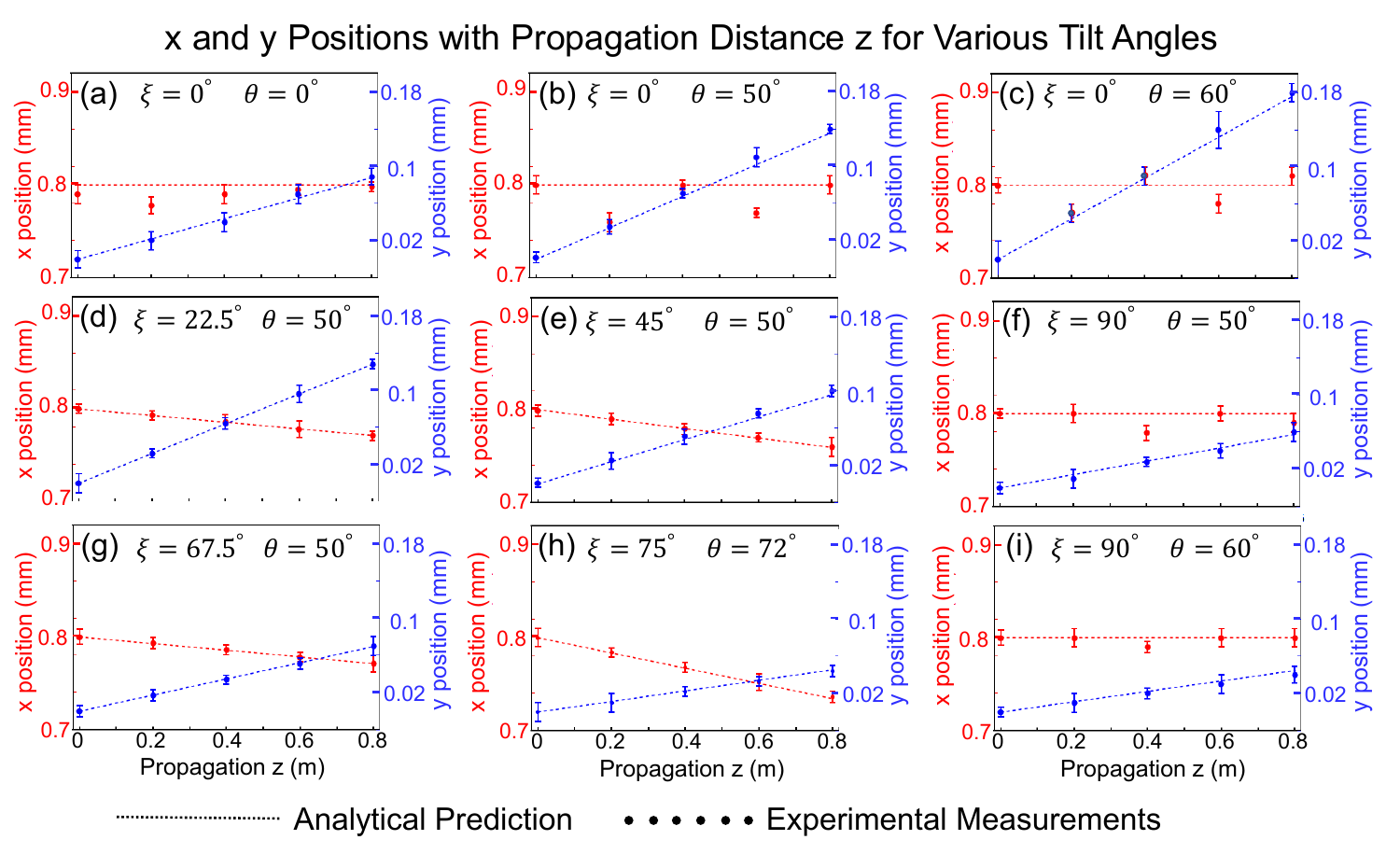}
\caption{\textit{Experimentally Measured and Analytically Predicted Vortex Trajectories.}  Experimental measurements of vortex position as a function of propagation for both $x$ and $y$ directions for various tilt angles. The measured data are points with marked error bars. Dashed lines show the analytical prediction from Eq. (22) in the manuscript. These results show that the tilt angle controls the vortex speed.
\label{fig:singlevortexpositions}}
\end{figure}

%%%%%%%%%%%%%
To experimentally measure vortex velocities reported in Fig. 5 of the manuscript, the vortex location at each $z$-step is plotted and fitted with a line. The error bars show the standard deviation across five measurements for each position. Examples of the measured vortex trajectories and the analytical prediction from Eq. (22) of the manuscript are shown here in Fig. \ref{fig:singlevortexpositions}. Dashed lines in Fig. \ref{fig:singlevortexpositions} are not linear fits to the data, but analytical predictions of Eq. (22) in the paper. For each tilt, linear fits weighted by the individual uncertainties\cite{taylor1997error} are separately calculated for the $x$ and $y$ trajectory. The results from these fits are used to plot the velocities shown with crosshairs bounded by ellipses denoting the uncertainty in each direction for Fig. 5 of the manuscript.

%%%%%%%%%%%%%
\subsubsection{Error Analysis of Vortex Trajectory Fits for Vortex Velocity Measurements}
%%%%%%%%%%%%%

Our vortex location measurements of ($x,y$) motion as a function of $z$ are each a set of measurements ($x_i,y_i$) with unique uncertainties ($\sigma_{xi},\sigma_{yi}$). We want to determine the velocity by separately fitting the $x$ and $y$ data to  $q=A + Bz$, where $B$ is the velocity and $q$ is either the $x$ or $y$ direction. Following the error analysis in ``An Introduction to Error Analysis'' by John. R. Taylor on pg. 198 \cite{taylor1997error}, we see that the propagating the errors results in a best fit slope $B$, and uncertainty in slope $\sigma_B$, of
\begin{equation}
    B = \frac{\sum w \sum w z q - \sum w z \sum w q}{\sum w \sum w z^2 -(\sum wz)^2},
\end{equation}
\begin{equation}
    \sigma_B=\sqrt{\frac{\sum w}{\sum w \sum wz^2-(\sum wz)^2}}
\end{equation}
\noindent where $w_i=1/\sigma_i^2$. We use these formulas to calculate the slopes and the uncertainty in the slopes. The slope fits for a given $x$ and $y$ set are the center of a data point in Fig. 5; the uncertainty in each slope determines the size of the ellipse bounding the crosshair for that data point. We do not find the intercept as we don't need it to measure the velocity. 

%As an example, we consider the x-positions for the case of a vortex with tilt of $\xi=^\circ$ and $\theta=^\circ$.

%...??

%%%%%%%%%%%%%
\subsubsection{Tilt Evolution Measurement for Single Vortex Case}
%%%%%%%%%%%%%

\begin{figure}[h!]
\includegraphics[width=\textwidth]{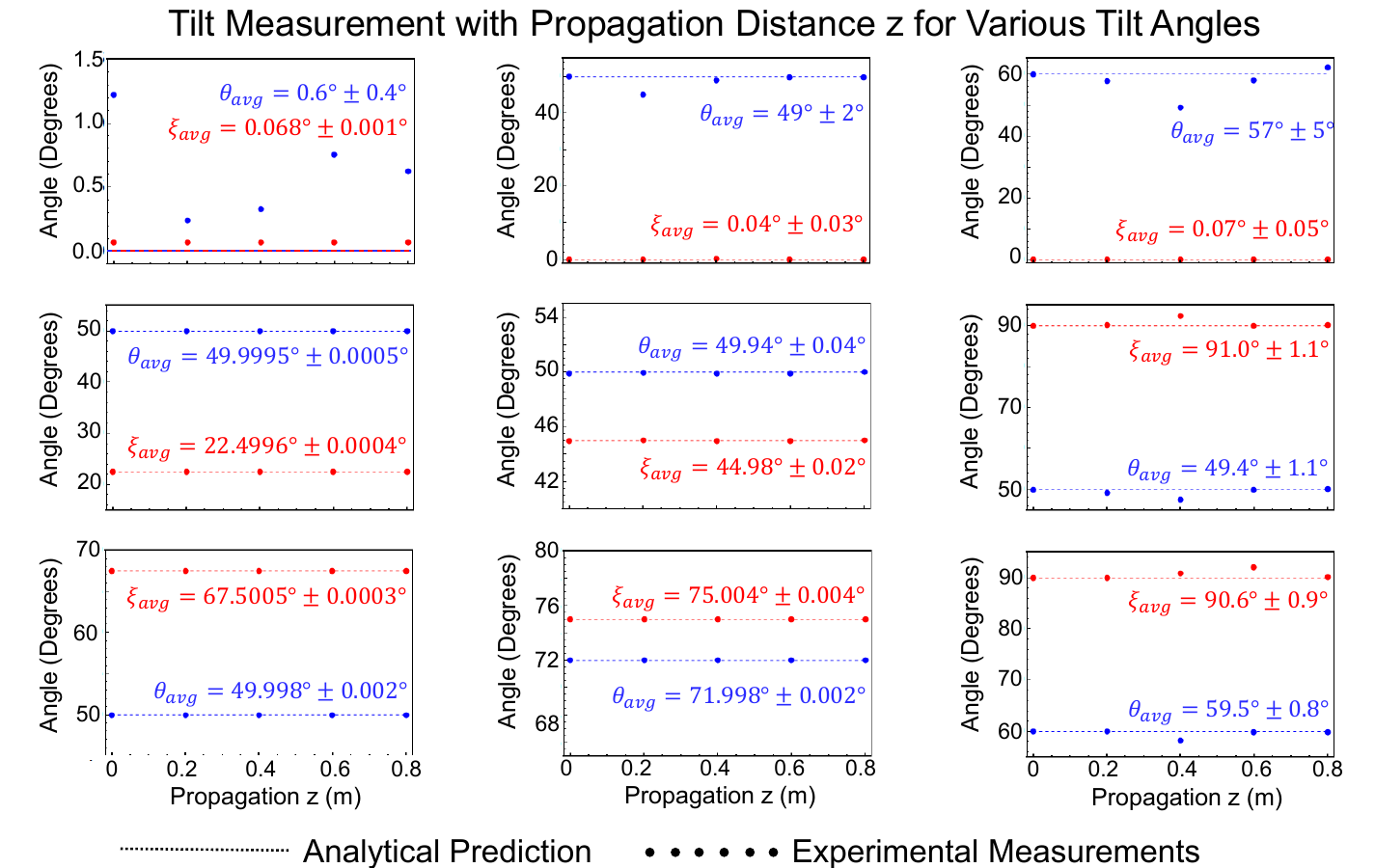}
\caption{\textit{Experimentally Measured and Analytically Predicted Vortex Tilt with Propagation.} The experimentally measured tilt values at each propagation step are shown for $\xi$ (red) and $\theta$ (blue) corresponding to the measurements in Fig. \ref{fig:singlevortexpositions}. The labels are the average value and standard deviation of each set of measurements. The dashed lines are the analytically predicted values, and it is clear that the experimental values are very well aligned with the predictions.   
\label{fig:singlevortextilt}}
\end{figure}

In addition to our measurements of vortex location as a function of propagation described in the last section, we also measured dynamic vortex tilt. A set of tilt measurements for various initial vortex tilts are shown in Fig. \ref{fig:singlevortextilt}, where the 2D phase is fit using Eq. (\ref{Single}). The average value along with one standard deviation are shown for each set of measurements, and are in excellent agreement with the expected values. The experimental tilt measurements are constant throughout propagation for the single, off-center vortex, which confirms the theoretical prediction.  

\vspace{2in}
%%%%%%%%%%%%%%%%%%%%%%%%%%%%%%%%%%%
\subsection{Two Vortex Generation and Measurement}
%%%%%%%%%%%%%%%%%%%%%%%%%%%%%%%%%%%

%%%%%%%%
\subsubsection{Two Vortex Holograms} 
%%%%%%

\begin{figure}[h!]
\begin{center}
\includegraphics[width=0.5\textwidth]{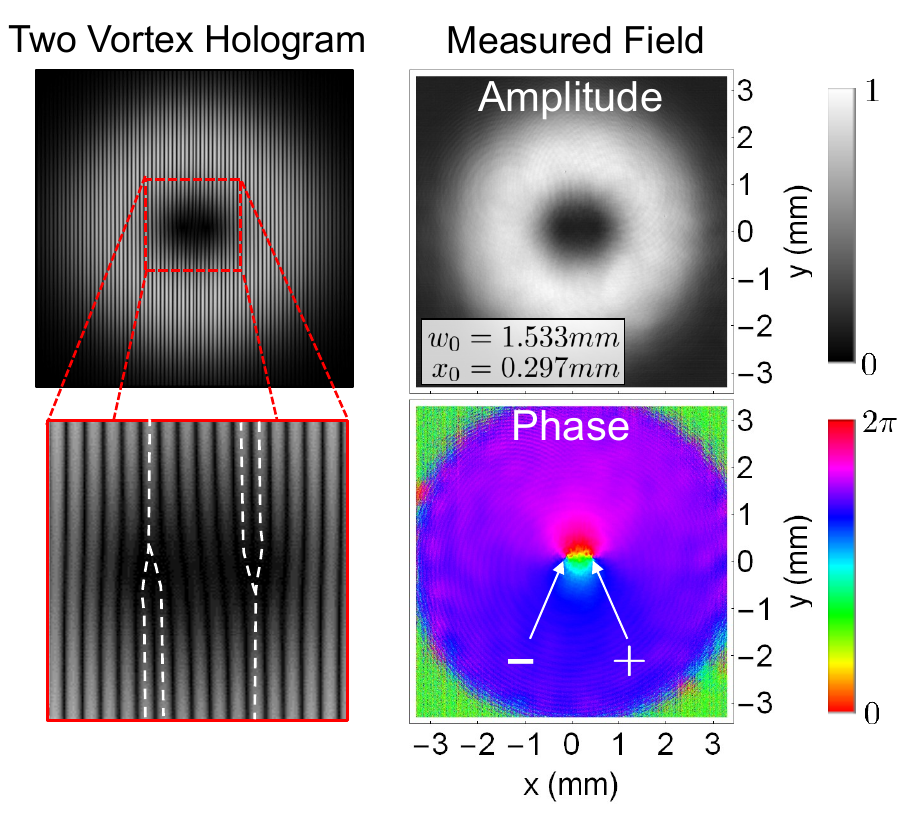}
\end{center}
\caption{ \emph{Vortex Pair Holograms and Measured $z=0$ mm Field.}  }
\label{fig:twovortexhologram}
\end{figure}

The hologram, generated using Eq. (\ref{eq:twovortex}), and the field measured at the $z=0$ imaging plane are shown in Fig. \ref{fig:twovortexhologram}. A zoomed in view of the vortex configuration and a white dashed outline traces over each fork to highlight the vortex locations in the hologram. The measured field amplitude and phase are shown in the right column. The inset on the amplitude plot shows the measured beam waist and vortex displacement, from a 2D amplitude fit using Eq. (\ref{eq:twovortex}).

%
%FIGURE 12 #
%
%%%% Experimental Data in Teams
% TFL -> Tilt (1- and 2-Vortex Repository -> Publication Data and Figures

%%%%%%%%
\subsubsection{Two Vortex Trajectories and y Velocity} 
%%%%%%

Zoomed in measurements of the amplitude and phase evolution, plotted in Fig. \ref{fig:twovortexampphase}, show the vortex annihilation dynamics. The vortices are marked with white circles in the phase, and move downward and towards each other until they annihilate each other and no vortices remain. Measurements following the annihilation event show no reappearance of the vortices.

%
%FIGURE #
%
\begin{figure}[h!]
\begin{center}
\includegraphics[width=0.6\textwidth]{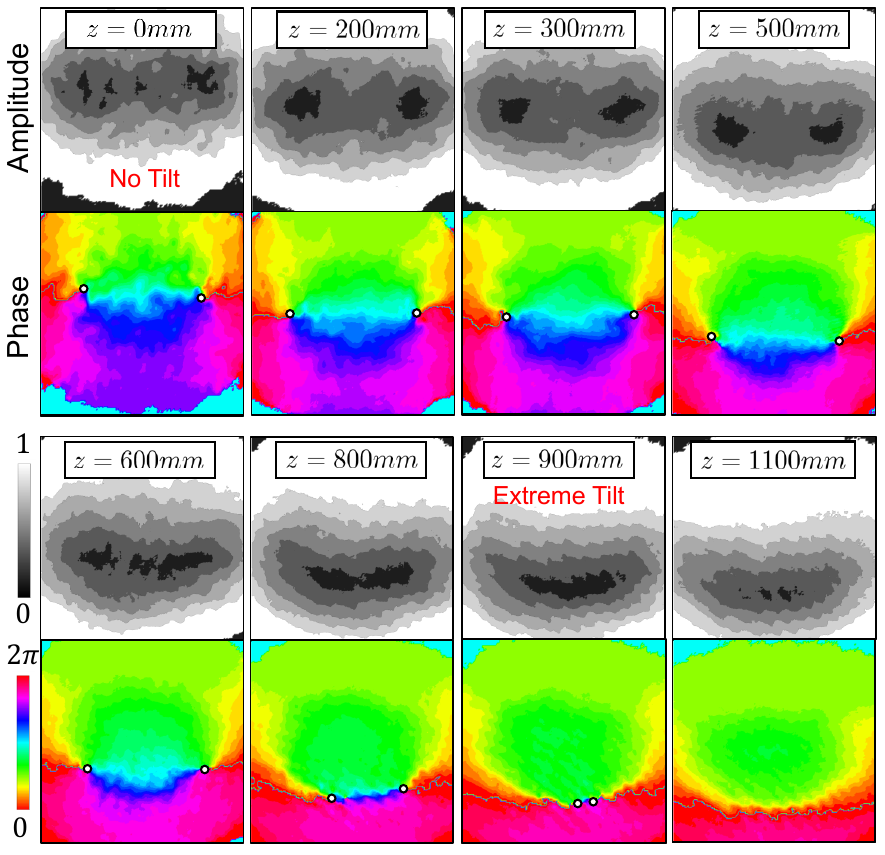}
\end{center}
\caption{ \emph{Experimental Amplitude and Phase Measurements.} Snapshots of the measured amplitude and phase with propagation show the vortex trajectory and annihilation event. Amplitude data are plotted with limited contours to highlight this evolution in the amplitude structure, but the data has equal resolution to that of the phase measurement. Data far beyond the annihilation point confirms that the vortices do not reemerge.
}
\label{fig:twovortexampphase}
\end{figure}

A qualitative look at the experimental amplitude measurements shows that the vortex tilt does evolve with propagation. In Fig. \ref{fig:twovortexampphase}, it is clear that the 2D amplitude structure of each vortex elongates and rotates with propagation.

From the tracked vortex locations, the $y$ velocity of the vortices is measured. Using a linear fit on the combined data for both the left and right vortex, the measured y velocity is $v_{y,meas}=0.389$ mm/m as compared to the predicted $v_{y,model}=0.314$ mm/m.

%
%FIGURE #
%
\begin{figure}[h!]
\begin{center}
\includegraphics[width=0.4\textwidth]{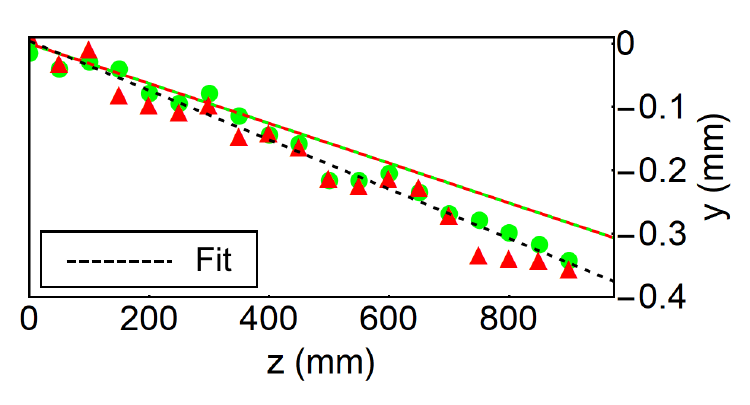}
\end{center}
\caption{ \emph{Experimental $y$ Velocity Measurment.} A constant velocity in the y direction is anticipated and measured. The fitted velocity (dashed line) shows that the measured velocity is slightly faster than the predicted velocity (green/red line).}
\label{fig:twovortexampphase}
\end{figure}

%%%%%%
\subsubsection{Two Vortex $V^2$ Matrix and Tilt Measurements} 
%%%%%%

From the tilt formalism in Sec. 2 of the Manuscript, we know that the gradient of $\psi$ is given by
$\nabla_\perp \psi = \nabla_\perp \textbf{F}^{-1} \textbf{r}_\perp = \textbf{F}^{-1}$. Using this relationship in conjunction with the relationship between $\Vsq$ and $\FLC$ from the Polar Decomposition Theorem, tilt angles can be found via
\begin{equation}
    \textbf{V}^2  = \left[ \nabla_\perp \psi^{-1} \right] \left[ \nabla_\perp \psi^{-1} \right]^T.
    \label{vsqexp}
\end{equation}
\noindent where
\begin{gather}
    \nabla_\perp \psi(x,y)=\nabla_\perp \begin{bmatrix} Re[\psi(x,y)] \\
    Im[\psi(x,y)]
    \end{bmatrix} =
    \begin{bmatrix} \partial_x Re[\psi(x,y)] & \partial_y Re[\psi(x,y)]  \\
    \partial_x Im[\psi(x,y)] &  \partial_y Im[\psi(x,y)]
    \end{bmatrix}.
    \label{eq:gradpsireim}
\end{gather}
This result shows that $\textbf{V}^2$ can be reconstructed in terms of the derivatives of the real and imaginary parts of the complex field at the location of the vortex for a given $z$ position -- i.e. the full measured field can be used and the background field does not have to be calculated.

\begin{figure}[h!]
\begin{center}
\includegraphics[width=0.75\textwidth]{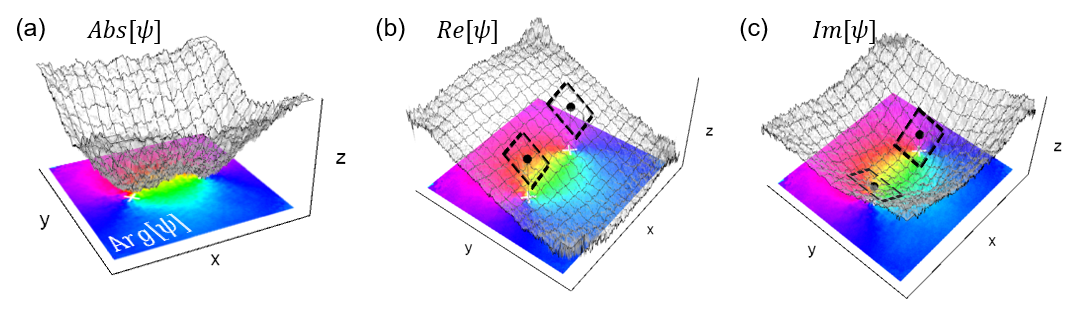}
\end{center}
\caption{ \textit{Visualization of Experimental Tilt Measurement for the Two Vortex Case:} (a) The complex field is shown with the absolute value of the field in a 3D mesh plot placed on top of the phase. (b) The vortex locations are marked at the real and imaginary zeros of $\psi$ with black dots and an example of fitted planes for each vortex over a chosen small window in the x-y plane is shown. The fit is given by $z_{re} = a_{re} x+ b_{re} y + c_{re}$. (c) the imaginary part of $\psi$, with a fit given by $z_{im} = a_{im} x+ b_{im} y + c_{im}$. }
\label{fig:twovortexgrad}
\end{figure}

As shown in Fig. \ref{fig:twovortexgrad}, to determine vortex location and tilt from experimental data, the measured complex field is separated into its real and imaginary parts with vortex locations in the $x$-$y$ plane found by the intersection of real and imaginary zeros. In calculating vortex tilt from vortex measurements, it is important to consider that $\textbf{F}$ still contains the background field (see Eq. (6) in the manuscript) while the eigenvalues and eigenvectors of $\textbf{V}^2$ are independent of both $\rho_{bg}$ and $\phi_{bg}$. With this as motivation, the $\textbf{V}^2$ matrix can be rewritten in terms of the gradient of an arbitrary field $\psi$.

To obtain the slope at each vortex location, a small window is first cropped around the selected vortex. The real and imaginary parts are separately fit to planes of the form $z = a x +b y+c$. These fit parameters, $a$ and $b$, for both real and imaginary parts are used to find $\xi$ and $\theta$, as shown in Fig. \ref{fig:twovortexgrad}. $\textbf{V}^2$ can be written in terms of these fit parameters as
\begin{gather}
    \textbf{V}^2=\frac{1}{(a_{re} b_{im}-a_{im} b_{re})^2} \begin{bmatrix} b_{im}^2+b_{re}^2 & -(a_{im}b_{im} + a_{re}b_{re})  \\
    -(a_{im}b_{im} + a_{re}b_{re}) & a_{im}^2+a_{re}^2
    \end{bmatrix}.
    \label{eq:gradpsireim}
\end{gather}

%\bibliography{Supplemental}